\documentclass[aps,a4paper, preprint, superscriptaddress,preprintnumbers,floatfix,nofootinbib,amsmath,amssymb]{revtex4}
\usepackage{url}
\usepackage{hyperref}
\usepackage{cleveref}

\usepackage{amstext,amssymb}
\usepackage[section]{placeins}
\usepackage{amsmath}
\usepackage{graphicx}
\usepackage{xspace}
\usepackage{color}
\usepackage{xcolor}
\usepackage{float}
\usepackage{comment}
\usepackage{amstext,amssymb}
\usepackage{amsmath}
\usepackage{graphicx}
\usepackage[section]{placeins}
\usepackage{xspace}
\usepackage{color}
\usepackage{units}
\usepackage{array}
\usepackage[section]{placeins}
\usepackage{amsmath,bm}
\usepackage{amssymb}
\usepackage{gensymb}
\usepackage{soul}
\usepackage{textcomp}
\usepackage{slashed} 
\usepackage{multirow}
\usepackage{hyperref}
\usepackage{mathrsfs}
\usepackage{enumitem}
\usepackage{graphicx}
\usepackage{color}
\usepackage{comment}
\usepackage{epstopdf}
\usepackage{float}
\usepackage{units}
\usepackage{slashed}
\usepackage{amsmath,bm}
\usepackage{slashed} 
\usepackage{multirow}

\def\s1{\hat s}

\usepackage{slashed}

\large
\begin{document}
\title{Probing Dual NSI and CP Violation in DUNE and T2HK}


\author{Barnali Brahma}
\email{ph19resch11001@iith.ac.in}

\author{Anjan Giri}
\email{giria@phy.iith.ac.in}

\affiliation{Department of Physics, IIT Hyderabad,
              Kandi - 502285, India }      




\begin{abstract}
The latest results from the long baseline neutrino experiments show a hint of non-zero CP violation in the neutrino sector. In this article, we study the CP violation effects in the upcoming long-baseline neutrino experiments DUNE and T2HK. Non-standard interactions can affect the cleaner determination of CP violation parameter. It has been argued that the NSI can help alleviate the tension between the recent $\delta_{CP}$ measurements of NO$\nu$A and T2K experiments. We consider here the dual NSI due to $\epsilon_{e\mu}$ and $\epsilon_{e\tau}$, arising simultaneously to see the effects in neutrino oscillation probabilities. Moreover, the CP asymmetry parameter $A_{CP}$ exhibits a clear distinction between normal and inverted mass orderings in the DUNE experiment.
\end{abstract}
\maketitle
\section{Introduction}
{Neutrino physics has come a long way since Pauli proposed the existence of neutrinos in 1930. The most elusive particles in nature, "Neutrinos", are present in abundance in the Universe. These neutral particles do not feel strong interactions and interact only via weak interactions. Neutrinos can have two types of interactions, they can couple with a $Z^{0}$ boson, or they can couple with a $W^{+}/W^{-}$ boson. A remarkable physical effect in the case of neutrinos, as originally predicted by Bruno Pontecorvo~\cite{Pontecorvo:1967fh}, is the existence of neutrino flavor oscillations. The success of  neutrino oscillation experiments~\cite{Super-Kamiokande:1998kpq, SNO:2002tuh, KamLAND:2004mhv, MINOS:2011neo} in the last few decades is one of the major achievements in modern physics. The results of several neutrino experiments employing solar~\cite{Davis:1968cp, Barger:1991ae, Miranda:2004nb}, atmospheric~\cite{Gonzalez-Garcia:1998ryc, Friedland:2005vy}, reactor, and accelerator neutrinos have played an important role in examining neutrino masses and mixing angles~\cite{Mohapatra:1979ia, Fogli:1996pv, Minakata:2001qm}. Six oscillation parameters determine the neutrino oscillation probabilities~\cite{Capozzi:2013csa} in the SM with three neutrinos:  three mixing angles ($\theta_{12},\theta_{13},\theta_{23}$), one Dirac CP phase $\delta_{CP}$, and two mass-squared differences ($\Delta m^{2}_{21}, \Delta m^{2}_{31}$). We have several shreds of evidence in the form of solar neutrinos and atmospheric neutrino anomalies to imply the occurrence of neutrino flavor change upon propagating long distances. Neutrino oscillation is an ideal avenue to probe new physics beyond the standard model. 

There are three important challenges: the value of  CP-violating phase $\delta_{CP}$, $\theta_{23}$ octant degeneracy, and the sign of mass-square difference ($\Delta m^{2}_{31}$), which are likely to be understood in the upcoming neutrino experiments. Of late, neutrino oscillation programs have been able to measure the unknown oscillation parameters with increasing accuracy~\cite{Forero:2014bxa, Esteban:2020cvm}. The effort of many dedicated neutrino oscillation experiments over the last two decades has started addressing the important questions of neutrino mass ordering~\cite{Nunokawa:2005nx}, octant degeneracy, and leptonic CP violation~\cite{Barger:1980jm, Arafune:1996bt, Donini:1999jc, Nunokawa:2007qh}. In this work, the framework of neutrino mass and the mixing concept are mostly discussed. Neutrinos interact with matter via weak SM interaction and gravity. 

The neutrinos, while traveling through the Earth's crust, get influenced by a matter potential known as Wolfenstein's matter effect~\cite{Wolfenstein:1977u, Guzzo:1991hi, Gonzalez-Garcia:2011vlg}. Wolfenstein introduced non-standard interaction (NSI)~\cite{Grossman:1995wx, Huber:2001zw, Fornengo:2001pm, Blennow:2007pu, Palazzo:2009rb, Coloma:2011rq, Esmaili:2013fva, MINOS:2013hmj, Choubey:2015xha, Miranda:2015dra, Bakhti:2016prn, Babu:2019vff, Majhi:2022wyp, Proceedings:2019qno, Ohlsson:2012kf, Farzan:2017xzy, Masud:2016bvp, Deepthi:2017gxg, Huitu:2016bmb, Denton:2022pxt} besides the neutrino mass matrix to probe new physics. Neutrino phenomenology has been extensively studied in the literature. NSI describes the non-standard neutrino interaction with ordinary matter, which is parameterized by effective couplings, $\epsilon_{\alpha \beta}$, where $\alpha$ and $\beta$ are the generation indices. In neutrino mass models, NSI arises while trying to explain the large as well as  small neutrino mixing angles~\cite{Farzan:2015hkd, Farzan:2018gtr, Biggio:2009nt}. The standard neutrino oscillation in the matter can get affected by NSI, and determining these subdominant effects will be the major role of the upcoming long-baseline neutrino experiments.

 Recently, T2K and NO$\nu$A  announced their results on neutrino CP-violating parameters. While for the inverted mass ordering, they look consistent with each other, but for the normal mass ordering, there appears to be some tension (T2K points to the $\delta_{CP}$ about 1.5$\pi$ whereas NO$\nu$A suggests the same parameter to be around 0.8$\pi$)~\cite{t2k, nova}. It will be intriguing to learn more about neutrino mass ordering from the results of the ongoing experiments T2K-NO$\nu$A, and it is expected to be a persistent issue in neutrino physics that will be resolved in the DUNE-T2HK era. Interestingly, attempts have been made to link the difference in $\delta_{CP}$, observed in T2K and NO$\nu$A, to the possibility of new physics in the form of non-standard interactions~\cite{Chatterjee:2020kkm, Denton:2020uda, Brahma:2022xld}. 

 In this work, we attempt to understand the CP-violation through the CP asymmetry parameter in the presence of vacuum, matter effects as well as in the presence of non-standard neutrino interactions arising simultaneously from $\epsilon_{e\mu}$ and $\epsilon_{e\tau}$ sectors, by considering the results of T2K-NO$\nu$A~\cite{T2K:2021xwb, NOvA:2021nfi}. Here, we focus on utilising the NSI constraints obtained by scanning two different NSI coupling parameters $\epsilon_{e\mu}$ and $\epsilon_{e\tau}$ at a time~\cite{Brahma:2023wlf}, which have not been discussed before. The possible implications of CP violation on neutrino mass ordering are discussed using the CP asymmetry parameter. The results shown in this analysis are for DUNE and T2HK experimental setup. The outline of the paper is as follows: first, we briefly discuss the formalism used in this analysis. Then we discuss the analysis details and results and observe the CP violation signatures. In the end, we present the summary and conclusions inferred from this analysis.
\section{Formalism}

The leptonic (PMNS) mixing matrix \textbf{U} is conventionally written in terms of the mixing angles $\theta_{12}$, $\theta_{13}$ and $\theta_{23}$ and of the
CP-violating phase $\delta_{CP}$ that plays a role in neutrino oscillations. The most common convention is,

\begin{equation*}
U_{PMNS} = 
\begin{bmatrix}
 c_{12}c_{13} & s_{12}c_{13} & s_{13} e{^{-i\delta_{cp}}} \\
-s_{12}c_{23}-c_{12}s_{13}s_{23}e^{i\delta_{cp}} & c_{12}c_{23}-s_{12}s_{13}s_{23}e^{i\delta_{cp}} & c_{13}s_{23}\\
s_{12}s_{23}-c_{12}s_{13}c_{23}e^{i\delta_{cp}} & -c_{12}s_{23}-s_{12}s_{13}c_{23}e^{i\delta_{cp}} & c_{13}c_{23}\\
\end{bmatrix}
\end{equation*}

where $s_{ij} , c_{ij} \equiv$ sin $\theta_{ij}$ , cos $\theta_{ij}$ and  the standard model(SM) phase $\delta_{cp} \in [0, 2\pi) $

\subsection{Neutrino Oscillation in Vacuum}
Vacuum oscillations are caused by the relative phases that the components of a neutrino with a given flavor obtain in the course of the time t. 
Using the above mixing matrix $U_{PMNS}$, the oscillation probabilities in the vacuum, when the flavor is unchanged (survival or disappearance probability),
\begin{equation}
    P_{\nu_{\alpha}\rightarrow \nu_{\alpha}} = 1-\sum_{i>j}4|U_{li}^{2}U_{lj}^{2}| sin^{2}\big(\Delta m^{2}_{ij}\frac{L}{4E}\big)
\end{equation}
or when the flavor of the neutrino changes (appearance probability).
\begin{equation}
    P_{\nu_{\alpha}\rightarrow \nu_{\beta}} = -4\sum_{i>j} \mathbb{R}\big(U_{\alpha i}^{*} U_{\beta i} U_{\alpha j} U_{\alpha j}^{*}\big) sin^{2}\big(\Delta m^{2}_{ij}\frac{L}{4E}\big)+2\sum_{i>j}\mathbb{I}\big(U_{\alpha i}^{*} U_{\beta i} U_{\alpha j} U_{\alpha j}^{*}\big) sin\big(\Delta m^{2}_{ij}\frac{L}{2E}\big)
\end{equation}

Where $\alpha,\beta$ are the neutrino flavors, $\Delta m^{2}_{ij}\equiv m_{i}^{2}-m_{j}^{2}$,  L is in meters and E in MeV (or L is in km and E in GeV).

\subsection{Neutrino oscillation in Matter}

The propagation of electron neutrinos in a material medium receives a special, additional phase of scattering, as discussed by Wolfenstein. On entering the Earth’s atmosphere and traveling through the Earth’s crust, the neutrino gets influenced by a matter potential known as Wolfenstein’s matter effect. The neutrino propagation Hamiltonian in the presence of matter, NSI, can be expressed as

\begin{align*}
H_{Eff} = \frac{1}{2E}
\Bigg[ U_{PMNS}
\begin{bmatrix}
 0 & 0 & 0 \\
0 & \Delta{m^{2}_{21}} & 0\\
0 &0 & \Delta{m^{2}_{31}} \\
\end{bmatrix}
 U^{\dagger}_{PMNS}
 + V\Bigg]
\end{align*}
where unitary Potecorvo-Maki-Nakagawa-Sakata mixing matrix is denoted by $U_{PMNS}$, neutrino energy as E, the different mass eigenstates as $m_{1}$, $m_{2}$ and $m_{3}$ and $\Delta m_{21}^{2}\equiv\ m_{2}^{2}-m_{1}^{2}$, $\Delta m_{31}^{2}\equiv m_{3}^{2}-m_{1}^{2}$. $V$ is written as:

\begin{equation*}
V = 2\sqrt{2}G_{F}N_{e}E
\end{equation*}
$N_{e}$ is the number density of electrons for neutrino propagation in the Earth. The mixing matrix in the matter, $U_{PMNS}$, is given by the following change of the parameters from the vacuum solution:

\begin{equation*}
    \sin^{2} 2\theta_{m} = \frac{\sin^{2} 2\theta}{\sin^{2} 2\theta+(\cos 2\theta - \xi)^{2}}
\end{equation*}

\begin{equation*}
    (\Delta m^{2})_{eff} = \Delta m^{2} \times \sqrt{\sin^{2} 2\theta + (\cos 2\theta - \xi)^{2}}
\end{equation*}

where  $\xi = \frac{2\sqrt{2}G_{F}N_{e}}{\Delta m^{2}}$, $\Delta m^{2} = \Delta m^{2}_{ij}\frac{L}{4E}$, $G_{F}$ is Fermi coupling constant, and $N_{e}$ is the number density of electrons.
The oscillation probabilities $P_{\nu_{\alpha} \rightarrow \mu_{\beta}}$ ($\alpha, \beta = e, \mu, \tau$ ) have the same forms as for the vacuum oscillations with mass
eigenstates as above and with replacements $\theta \rightarrow \theta_{m}$. 

\subsection{Neutrino oscillation in the presence of NSIs}

The NSI is defined by dimension six four-fermion ($ff$) operators of the form~\cite{Wolfenstein:1977u}
\begin{equation}\label{1}
    {\mathcal{L}}_{NSI} = 2\sqrt{2}G_{F} \epsilon_{\alpha \beta}^{fC} [  \overline{\nu_{\alpha}} \gamma^{\rho} P_{L}  \nu_{\beta}][\overline{f} \gamma_{\rho} P_{C} f] + h.c.
\end{equation}
In the above expression, $\epsilon_{\alpha \beta}^{fC}$ is the dimensionless parameters that assess the strength of the new interaction with respect to the SM, the three SM neutrino flavor $e, \mu, \tau$ are denoted by $\alpha$ and  $\beta$, matter fermions $f$ by  $u, d, e $, and the superscript $C = L, R$ refers to the chirality of $ff$ current. 
In the presence of matter, NSI, the Hamiltonian for neutrino propagation, can be expressed as,

\begin{align*}
H_{Eff} = \frac{1}{2E}
\Bigg[ U_{PMNS}
\begin{bmatrix}
 0 & 0 & 0 \\
0 & \Delta{m^{2}_{21}} & 0\\
0 &0 & \Delta{m^{2}_{31}} \\
\end{bmatrix}
 U^{\dagger}_{PMNS}
 + V\Bigg]
\end{align*}
Here, we have denoted neutrino energy as E, the different mass eigenstates as $m_{1}$, $m_{2}$, and $m_{3}$. The mass square differences are, $\Delta m_{21}^{2}\equiv\ m_{2}^{2}-m_{1}^{2}$, $\Delta m_{31}^{2}\equiv m_{3}^{2}-m_{1}^{2}$. The matter potential $V$ in the presence of NSI is written as:

\begin{equation*}
V = 2\sqrt{2}G_{F}N_{e}E
\begin{bmatrix}
  1+\epsilon_{ee}& \epsilon_{e \mu}e^{i \phi_{e \mu}} & \epsilon_{e \tau}e^{i \phi_{e \tau}} \\
  \epsilon_{ \mu e}e^{-i \phi_{e \mu}}  & \epsilon_{\mu \mu} & \epsilon_{\mu \tau}e^{i \phi_{\mu \tau}} \\
  \epsilon_{\tau e} e^{-i \phi_{e \tau}} & \epsilon_{\tau \mu}e^{-i \phi_{\mu \tau}} & \epsilon_{\tau \tau}\nonumber\\
\end{bmatrix}
\end{equation*}

For neutrino propagation in the Earth, the NSI coupling:\\ \begin{center}
    $\epsilon_{\alpha\beta}e^{i\phi_{\alpha \beta}} \equiv \sum_{f, C}\epsilon_{\alpha\beta}^{fC} \frac{N_{f}}{N_{e}} \equiv \sum_{f=e,u,d}(\epsilon_{\alpha\beta}^{fL}+\epsilon_{\alpha\beta}^{fR}) \frac{N_{f}}{N_{e}}$,
\end{center}
$N_f$ being the number density of $f$ fermion and $N_{e}$ the number density of electrons. For this analysis, we utilise flavor non-diagonal NSI ($\epsilon_{\alpha \beta}$'s with $\alpha \neq \beta$). Mainly, we focus on the dual NSI parameter $\epsilon_{e \mu}$ and $\epsilon_{e \tau}$ (arising simultaneously) to examine the conversion probability of $\nu_{\mu}$ to $\nu_{e}$. For dual NSI scenario, the oscillation probability expression for $\nu_{\mu} \rightarrow \nu_{e}$ can be written as the sum of four (plus higher order; cubic and beyond) terms~\cite{Liao:2016hsa, Kikuchi:2008vq, Kopp:2007ne, Meloni:2009ia, Brahma:2023wlf}: 
\begin{equation}
P_{\mu e} = P_{SM} + P_{\epsilon_{e\mu}} + P_{\epsilon_{e\tau}} + P_{Int} + h.o.
\end{equation}
the above Eq.(4) can be explicitly written as:

\begin{equation*}
P_{SM} = 4[s_{13}^{2}s_{23}^{2}f^{2}+r^{2}s_{12}^{2}c_{12}^{2}c_{23}^{2}g^{2}]+8s_{13}s_{23}s_{12}c_{12}c_{23}rfg\cos({\Delta+\delta_{CP}})
\end{equation*}
\begin{eqnarray}
P_{\epsilon_{e\mu}}&=& 4\hat{A}\epsilon_{e\mu} [xf^{2}s^{2}_{23}\cos({\Psi_{e\mu}})+xfgc^{2}_{23}\cos({\Delta+\Psi_{e\mu}}) + yg^{2}c^{2}_{23}\cos{\phi_{e\mu}} \nonumber\\
&+& ygfs^{2}_{23}\cos({\Delta - \phi_{e\mu}})]
 +4\hat{A}^{2}\epsilon_{e\mu}^{2}[g^{2}c^{4}_{23}+f^{2}s_{23}^{4}+2fgs_{23}^{2}c^{2}_{23}\cos{\Delta}]\nonumber
 \end{eqnarray}
 \begin{eqnarray}
 P_{\epsilon_{e\tau}}&=& 4\hat{A}\epsilon_{e\tau}s_{23}c_{23}[xf^{2}\cos({\Psi_{e\tau}})-xfg\cos({\Delta+\Psi_{e\tau}}) - yg^{2}\cos{\phi_{e\tau}} \nonumber\\
&+& ygf\cos({\Delta - \phi_{e\tau}})]
 +4\hat{A}^{2}\epsilon_{e\tau}^{2}c^{2}_{23}s^{2}_{23}[g^{2}+f^{2}-2fg\cos{\Delta}]\nonumber
\end{eqnarray}
\begin{eqnarray}
P_{Int}&=& 8\hat{A}^{2}s_{23}c_{23}\epsilon_{e\mu}\epsilon_{e\tau}[g^{2}c^{2}_{23} +f^{2}s^{2}_{23}] + 8\hat{A}^{2}fgs_{23}c_{23}\epsilon_{e\mu}\epsilon_{e\tau}[2c^{2}_{23}\cos(\phi_{e\mu}-\phi_{e\tau})\cos \Delta\nonumber\\
    &-&\cos{(\Delta-\phi_{e\mu}+\phi_{e\tau})}] \nonumber
\end{eqnarray}
where, $x\equiv2s_{13}s_{23}$ and $y\equiv2rs_{12}c_{12}c_{23}$; $g \equiv \frac{\sin{\hat{A}\Delta}}{\hat{A}}$; ${f\equiv \frac{\sin{[(1-\hat{A})\Delta]}}{1-\hat{A}}}$;$\Delta\equiv |\frac{\Delta m^{2}_{31}L}{4E}|$; $r=|\frac{\Delta m^{2}_{21}}{\Delta m^{2}_{31}}|$; $\hat{A}\equiv|\frac{2\sqrt{2}G_{F}N_{e}E}{\Delta m^{2}_{31}}|$. Furthermore, here we use: $\Psi_{e\mu}=\phi_{e\mu}+\delta_{CP}$; $\Psi_{e\tau}=\phi_{e\tau}+\delta_{CP}$, where $\phi_{e\mu}$ and $\phi_{e\tau}$ are the non-standard CP-phases corresponding to the relevant NSI coupling $\epsilon_{e\mu}$ and $\epsilon_{e\tau}$ respectively. The dual NSI effects arising simultaneously from $\epsilon_{e\mu}$ and $\epsilon_{e\tau}$ are given in the term $P_{Int}$.\\

In the case for anti-neutrino probability, $P (\overline{\nu}{_{e}} \rightarrow \overline{\nu}{_{\mu}} )$, is given by
changing the above expression for $P_{SM}$, $P_{\epsilon_{e\mu}}$, $P_{\epsilon_{e\tau}}$, $P_{Int}$  with  $\delta \rightarrow -\delta$, $\phi_{\alpha\beta} \rightarrow -\phi_{\alpha\beta}$, and $\hat{A} \rightarrow -\hat{A}$ (and hence $f \rightarrow \overline{f}$), and for the inverted hierarchy (IH), $\delta_{CP} \rightarrow -\delta_{CP}$, $\hat{A} \rightarrow -\hat{A}$ (i.e., $f \leftrightarrow -\overline{f}$, and $g \leftrightarrow -g$) and $y \rightarrow -y$.

Using the oscillation probability, we can try to access the CP asymmetry observable, which in turn will help us to understand the CP violation. The CP asymmetry is the difference between the quantity of matter and anti-matter in the universe. The CP asymmetry measures the change in oscillation probabilities when the sign of the CP phase changes. CP-asymmetry is defined as:
\begin{equation}
    A_{CP} \equiv \frac{P(\nu_{\alpha}\rightarrow\nu_{\beta})-P(\overline{\nu}_{\alpha}\rightarrow\overline{\nu}_{\beta})}{P(\nu_{\alpha}\rightarrow\nu_{\beta})+P(\overline{\nu}_{\alpha}\rightarrow\overline{\nu}_{\beta})},
\end{equation}

Here, along with $A_{CP}$ observable, we use another observable for our illustration.
\begin{equation}
    \Delta A_{\alpha \beta}(\delta_{CP}) \equiv A_{\alpha \beta}(\delta_{CP}\neq0) - A_{\alpha \beta}(\delta_{CP}=0),
\end{equation}  

We briefly discussed the CP violation effects in the presence of single NSI coupling arising either from $\epsilon_{e\mu}$ and $\epsilon_{e\tau}$ \cite{fpcp2023}, where we see a clear distinction between normal and inverted mass hierarchies. In this work, we make an elaborate study of CP asymmetries and also see appreciable differences but now with non-standard interactions arising from $\epsilon_{e\mu}$ and $\epsilon_{e\tau}$ sectors simultaneously or dual NSIs. 

\section{Analysis Details}
For our analysis purpose, we use the software GLoBES~\cite{GLoBES, Huber:2004ka, Huber:2007ji} and its supplementary public tool, which helps us to utilise the non-standard interactions in our analysis. The standard model parameters' best-fit values are taken from NuFIT v5.2~\cite{nufit}. In particular, the parameter values taken for our analysis (for normal ordering) are: $\sin^{2}\theta_{12}=  0.303^{+0.012}_{-0.012}$;\hspace*{0.1cm}$\sin^{2}\theta_{13}= 0.02225^{+0.00056}_{-0.00059}$;\hspace*{0.1cm}
$\sin^{2}\theta_{23}=0.451^{+0.019}_{-0.016}$;\hspace*{0.1cm} $\delta_{CP}=232^{+36}_{-26}$;
$\frac{\Delta m^{2}_{21}}{10^{-5}eV^{2}}= 7.41^{+0.21}_{-0.20}$; and $\frac{\Delta m^{2}_{3l}}{10^{-3} eV^{2}}=+2.507^{+0.026}_{-0.027}$. As mentioned earlier, we assume that the new physics effects are responsible for different values of $\delta_{CP}$, and in this case, it is due to the presence of NSIs. The new physics parameters are in the form of non-standard interaction couplings, $\epsilon_{e\mu}$ and $\epsilon_{e\tau}$ arising simultaneously, which is first discussed in Ref.~\cite{Brahma:2023wlf}, and are summarized in Table I. The constraints obtained are consistent with the existing bounds given in Ref.~\cite{Coloma:2019mbs, IceCubeCollaboration:2021euf}. However, in this work, we focus on the CP asymmetry studies, which are not discussed in Ref.~\cite{Brahma:2023wlf}. Using the experimental setup of DUNE and T2HK, we study the leptonic CP-violating effects using the SM framework with matter effects and dual NSI setup.

In the following subsections, we discuss oscillation probability and $A_{CP}$ parameter individually for both DUNE and T2HK experimental setup, and in the last subsection, we exhibit a comparison between both DUNE and T2HK to understand the difference. 

\subsection{DUNE}
Here, we use DUNE~\cite{DUNE:2020ypp} running for 3.5 years and 3.5 years in $\nu$ and in $\Bar{\nu}$ mode, respectively. DUNE will have a 40-kiloton liquid argon detector that will generate neutrino and antineutrino beams from in-flight pion decays using a 1.2 MW proton beam. Fermilab, 1300 km upstream, will be the source of the proton beam. As an off-axis experiment, the neutrinos' energy will have a flux peak at around 2.6 GeV.  

\begin{table}[h!]
\caption{\label{tab:table1}{The best-fit points are listed here. }}
\begin{center}
\begin{ruledtabular}
\begin{tabular}{ccc}
 \large{Mass ordering} &  $|\epsilon_{e \mu}|$ & $|\epsilon_{e \tau}|$  \\ [2ex]  \hline
NO  & 0.1  & 0.033   \\ 
IO  & 0.1 & 0.02 \\ [1ex]
\hline
 \large{Mass ordering} & $\phi_{e\mu}/\pi$ & $\phi_{e \tau}/\pi$  \\ [2ex]\hline
NO  & 1.06  & 1.87  \\
IO  & 1.0 & 1.73 \\ 
\end{tabular}
\end{ruledtabular}
\end{center}
\end{table}

In Figure 1, we show DUNE standard model oscillation probability plots versus energy for NO in vacuum for neutrino (top panel) and anti-neutrino (bottom panel) sectors. In the top right panel, we have neutrino oscillation probability versus energy plots for $\delta_{CP}=0$ and $\delta_{CP}=232^{\circ}$ (from nuFIT v5.2). On the left-hand side, we have the effective oscillation probability difference ($P_{\delta=232^{\circ}}$-$P_{\delta=0^{\circ}}$) plot.  In the case of the DUNE, we consider an energy window of 2 GeV to 3 GeV. We have restricted ourselves to the energy range around the peak neutrino beam for the sake of illustration. Here, the average positive probability difference in DUNE's energy window for neutrino is around 1.2$\%$. Similarly, in the case of anti-neutrino, the right side bottom plot gives the oscillation probability versus energy plot for $\delta_{CP}=0$ and $\delta_{CP}=232^{\circ}$. Here, the average negative probability difference in DUNE's energy window for anti-neutrino is around 0.6$\%$. Here, the average probability difference value quantifies the change in probability with a change in the $\delta_{CP}$ phase in vacuum.
\begin{figure}[htbp]
\minipage{0.50\textwidth}
   \includegraphics[width=8.0cm,height=8.0cm]{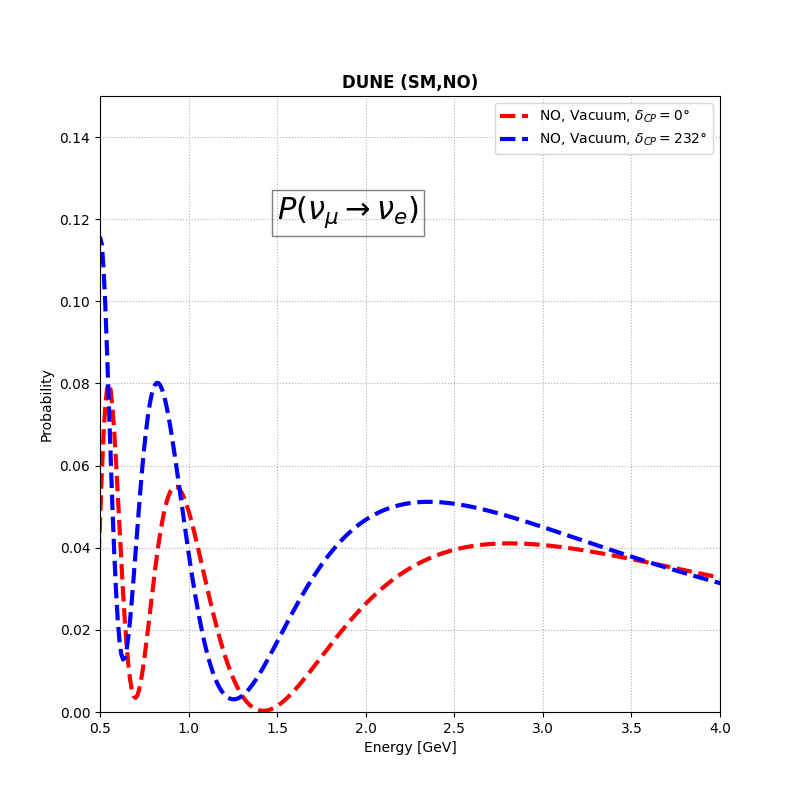}
\endminipage\hfill
\minipage{0.50\textwidth}
   \includegraphics[width=8.0cm,height=8.0cm]{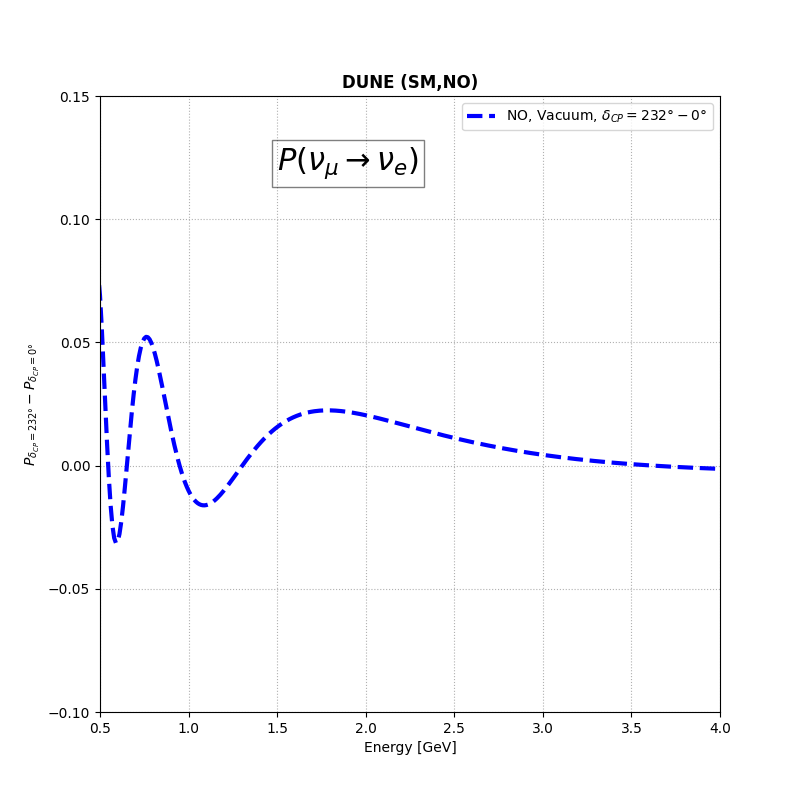}
\endminipage

\minipage{0.50\textwidth}
  \includegraphics[width=8.0cm,height=8.0cm]{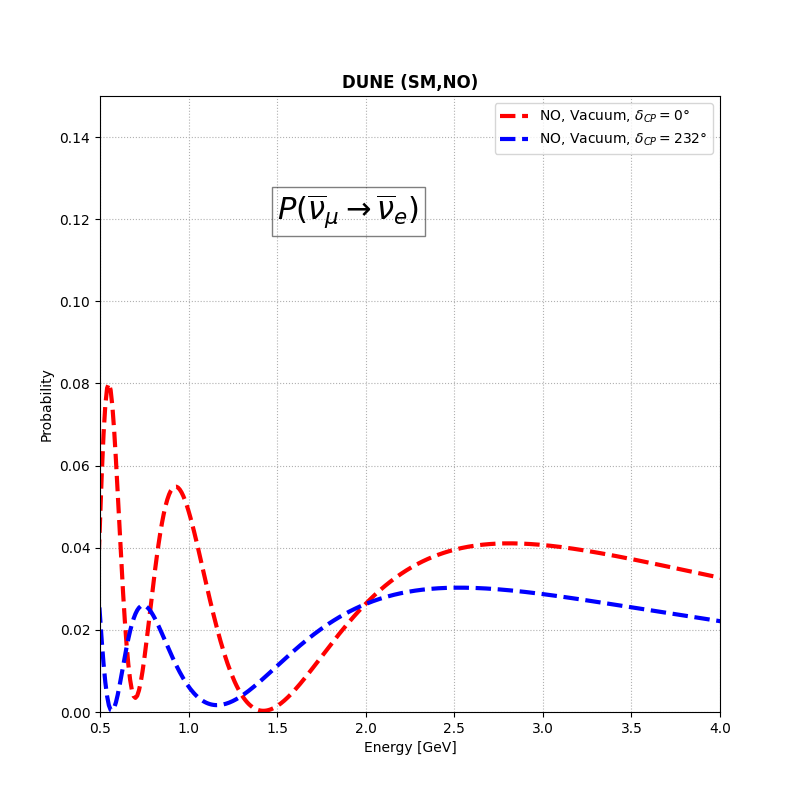}
\endminipage\hfill
\minipage{0.50\textwidth}
  \includegraphics[width=8.0cm,height=8.0cm]{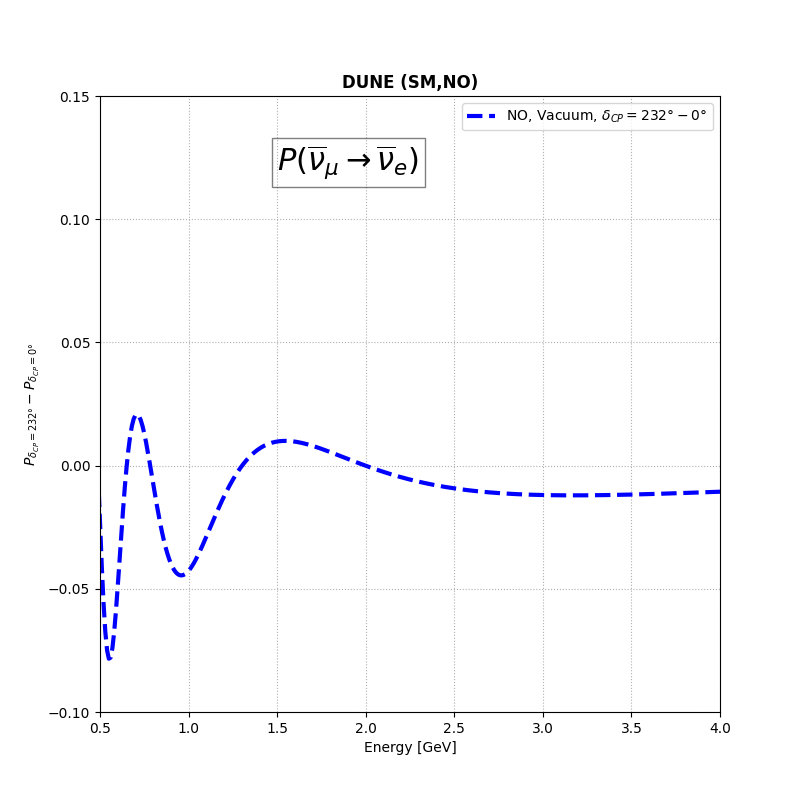}
\endminipage
\caption{Oscillation probability plots versus energy for NO in vacuum for neutrino (top panel) and anti-neutrino (bottom panel) sectors in DUNE experimental setup. Alongside oscillation  probability plots, we have oscillation probability difference plots (($P_{\delta=232^{\circ}}$-$P_{\delta=0^{\circ}}$) for neutrino (top right) and anti-neutrino (bottom right) }
\end{figure}

In Figure 2, we show oscillation probability plots in the presence of matter as well as vacuum for NO in the case of neutrino and anti-neutrino sectors in DUNE's experimental setup. The left side plots indicate oscillation probability versus energy ranging from 0 to 4 GeV, whereas the plots on the right depict the difference in oscillation probability between vacuum and matter. Using non-zero SM $\delta_{CP}=232^{\circ}$, as evident from the top left plot, the oscillation probability of the neutrino in the presence of matter is more dominating than in the case of vacuum. The top right plot quantifies that the average difference between $P_{Vacuum}$ and $P_{Matter}$ for the neutrino sector is around -2.2$\%$. Similarly, in the case of anti-neutrino (bottom), the oscillation probability in the vacuum is more dominating than the presence of matter. Here, the average difference between $P_{Vacuum}$ and $P_{Matter}$ for the anti-neutrino sector is around +1.4$\%$. In these plots, we visualise the change in oscillation probabilities when neutrinos travel through the vacuum and in the presence of matter medium.

\begin{figure}[htbp]
\minipage{0.50\textwidth}
   \includegraphics[width=8.0cm,height=8.0cm]{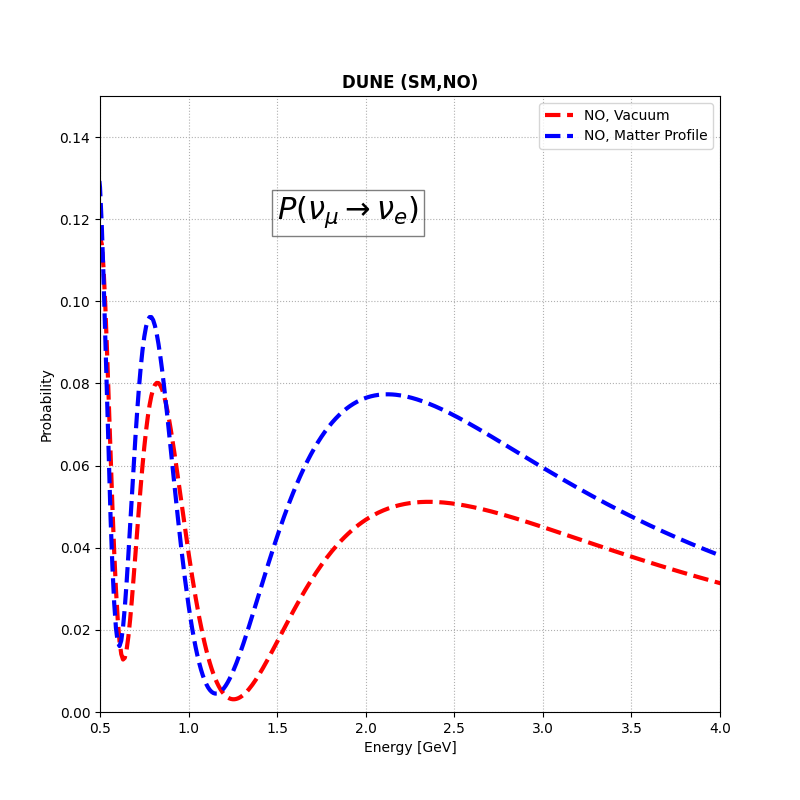}
\endminipage\hfill
\minipage{0.50\textwidth}
   \includegraphics[width=8.0cm,height=8.0cm]{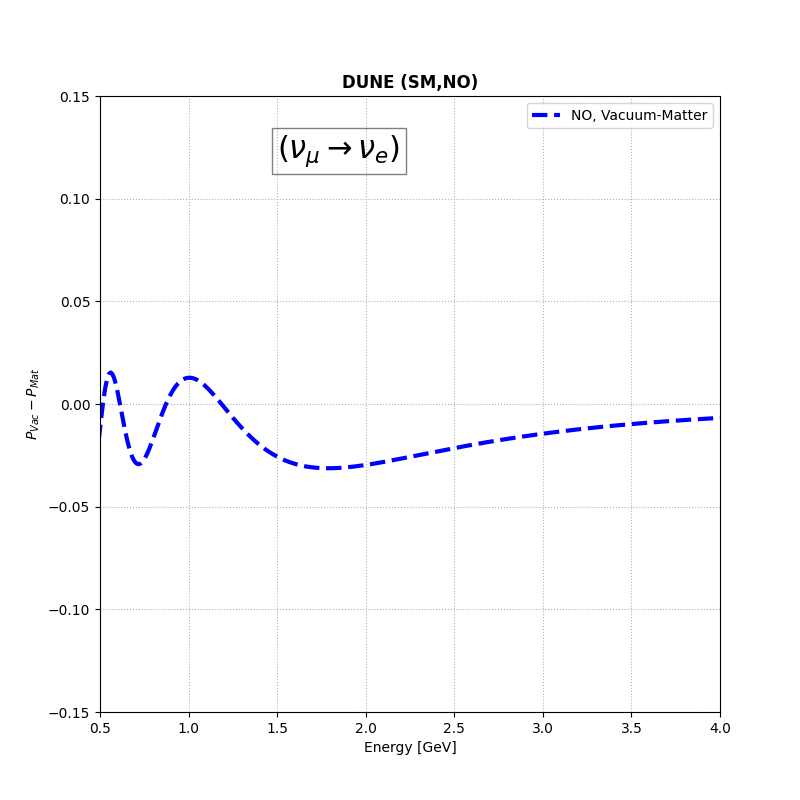}
\endminipage

\minipage{0.50\textwidth}
  \includegraphics[width=8.0cm,height=8.0cm]{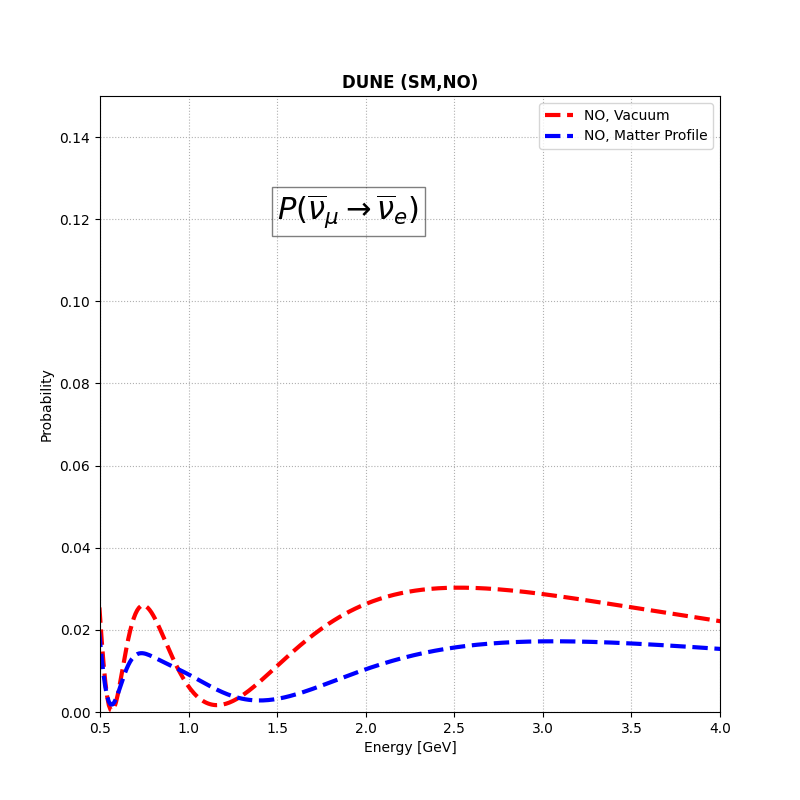}
\endminipage\hfill
\minipage{0.50\textwidth}
  \includegraphics[width=8.0cm,height=8.0cm]{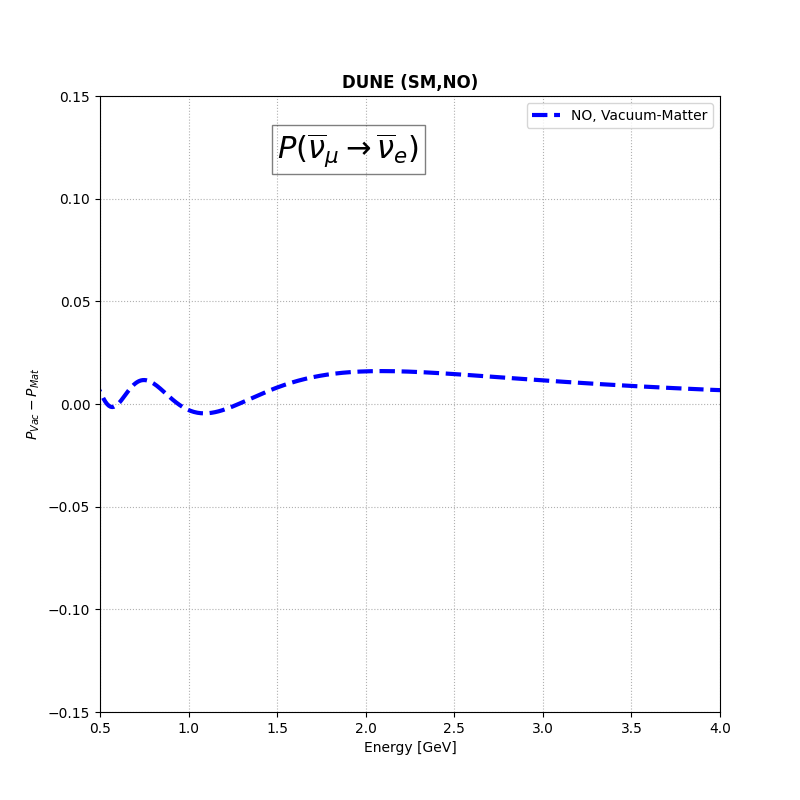}
\endminipage
\caption{Oscillation probability plots versus energy for NO in the vacuum and in the presence of matter for neutrino (top panel) and anti-neutrino (bottom panel) sectors in DUNE experimental setup. Alongside oscillation  probability plots, we have oscillation probability difference plots ($P_{Vacuum}$-$P_{Matter}$) for neutrino (top right) and anti-neutrino (bottom right) }
\end{figure}

In Figure 3, we show similar oscillation probability plots in the presence of matter as well as vacuum, but now for IO in the case of neutrino and anti-neutrino sectors in DUNE's experimental setup. The left side plots indicate oscillation probability versus energy ranging from 0 to 4 GeV, whereas the plots on the right depict the difference in oscillation probability between vacuum and matter. Using non-zero SM $\delta_{CP}=232^{\circ}$, as evident from the top left plot, the oscillation probability of the neutrino in the vacuum is more dominating than the oscillation probability in the presence of matter. The top right plot quantifies that the average difference between $P_{Vacuum}$ and $P_{Matter}$ for the neutrino sector is around 2.4$\%$. Similarly, in the case of anti-neutrino (bottom), the oscillation probability in the matter is more dominating than the vacuum. Here, the average difference between $P_{Vacuum}$ and $P_{Matter}$ for the anti-neutrino sector is around -2.2$\%$. Similar to the NO scenario, here, also  we visualise the change in oscillation probabilities when neutrinos travel through the vacuum and in the presence of matter medium.
\begin{figure}[htbp]
\minipage{0.50\textwidth}
   \includegraphics[width=8.0cm,height=8.0cm]{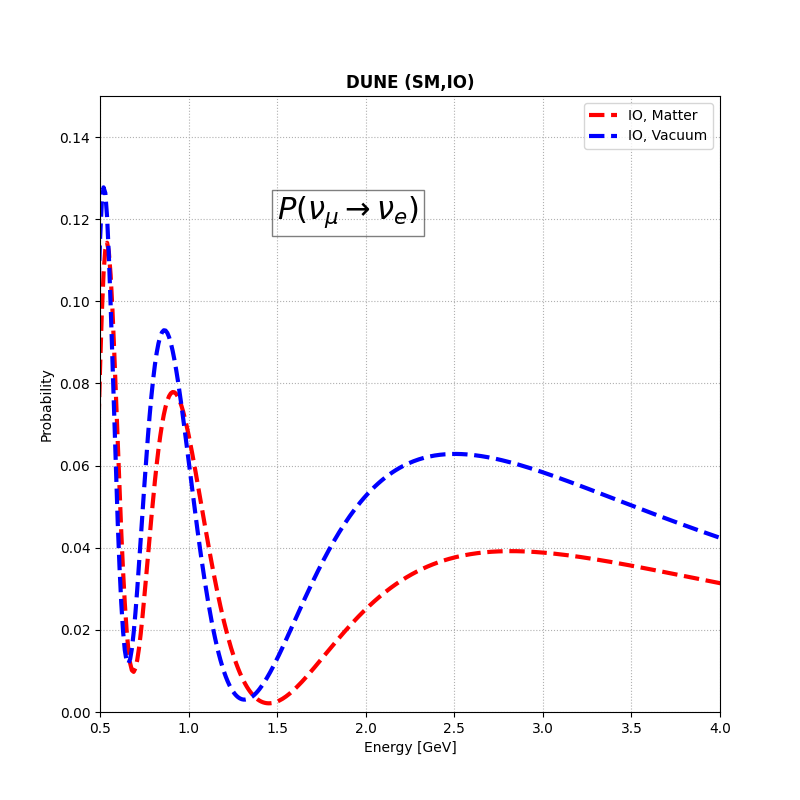}
\endminipage\hfill
\minipage{0.50\textwidth}
   \includegraphics[width=8.0cm,height=8.0cm]{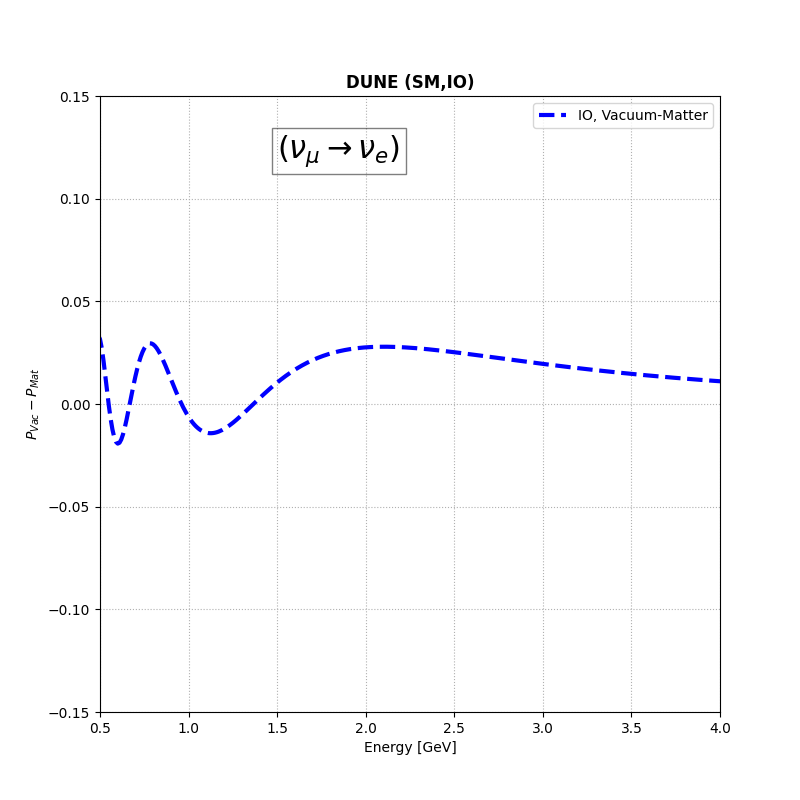}
\endminipage

\minipage{0.50\textwidth}
  \includegraphics[width=8.0cm,height=8.0cm]{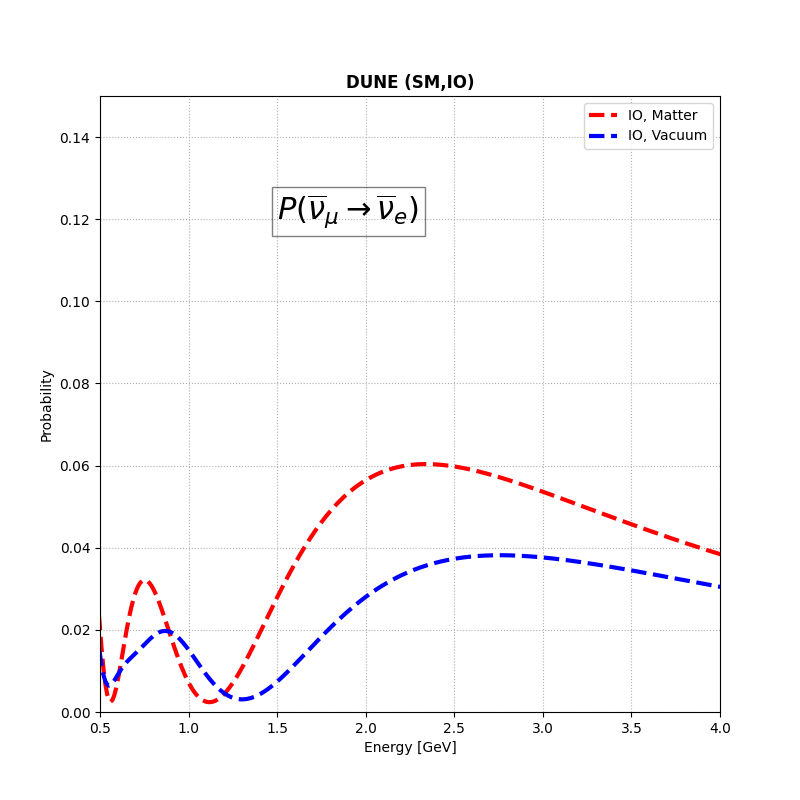}
\endminipage\hfill
\minipage{0.50\textwidth}
  \includegraphics[width=8.0cm,height=8.0cm]{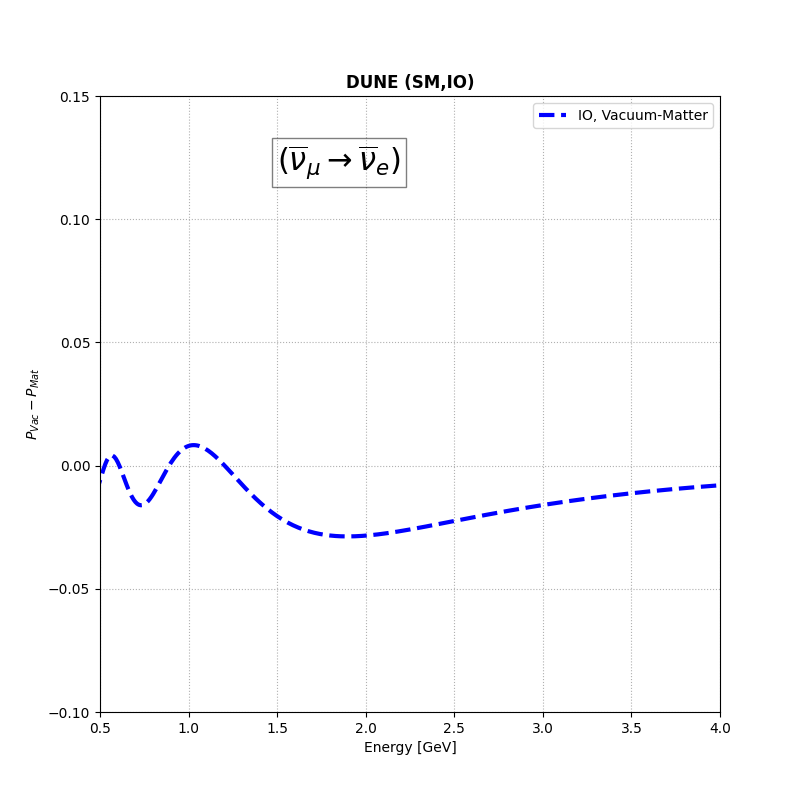}
\endminipage
\caption{Oscillation probability plots versus energy for IO in the vacuum and in the presence of matter for neutrino (top panel) and anti-neutrino (bottom panel) sectors in DUNE experimental setup. Alongside oscillation  probability plots, we have oscillation probability difference plots ($P_{Vacuum}$-$P_{Matter}$) for neutrino (top right) and anti-neutrino (bottom right)}
\end{figure}

In all the previous and subsequent plots for DUNE, we have considered the SM $\delta_{CP}$ to be 232$^{\circ}$ for NO and 276$^{\circ}$ for IO. In this particular figure, we have considered a nominal $\delta_{CP}$ value of 1.5$\pi$ for NO and IO. The motivation behind using a nominal value comes from the $\delta_{CP}$ discrepancy that arises in the case of T2K and NO$\nu$A. T2K has a shorter baseline in comparison to  NO$\nu$A, which will help us to determine the standard parameters almost independently of NSI. Thus, in Figure 4, we utilise the non-zero $\delta_{CP}$ value as given by T2K, i.e., 1.5$\pi$. In this figure, we show oscillation probability versus energy plots for neutrino NO (top left) as well as IO (top right) and also for anti-neutrino NO (bottom left) and IO (bottom right). In the neutrino NO case, the plot indicates that the oscillation probability in the presence of matter effect along with non-zero standard model $\delta_{CP}$ phase of 1.5$\pi$ is more dominating than the oscillation probability in the presence of matter effect with zero $\delta_{CP}$ phase. A similar dominating feature is also present in the neutrino IO case. For the anti-neutrino case, in the NO scenario, the oscillation probability in the presence of matter effect with zero $\delta_{CP}$ phase is more dominating than in the presence of matter effect with non-zero $\delta_{CP}$ phase of 1.5$\pi$. Again a similar feature is visualised in the plot for the anti-neutrino IO case. Here, we get the picture of a change in the oscillation probabilities with nominal 1.5$\pi$ and zero $\delta_{CP}$ phase in the presence of matter.
\begin{figure}[htbp]
\minipage{0.45\textwidth}
  \includegraphics[width=8.0cm,height=8.0cm]{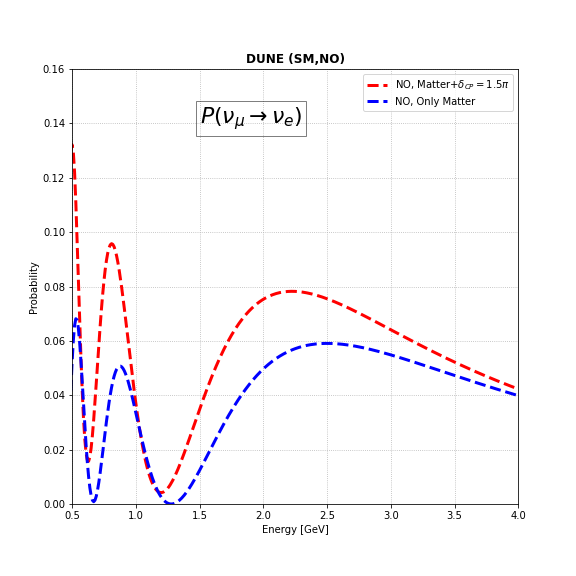}
\endminipage\hfill
\minipage{0.45\textwidth}
  \includegraphics[width=8.0cm,height=8.0cm]{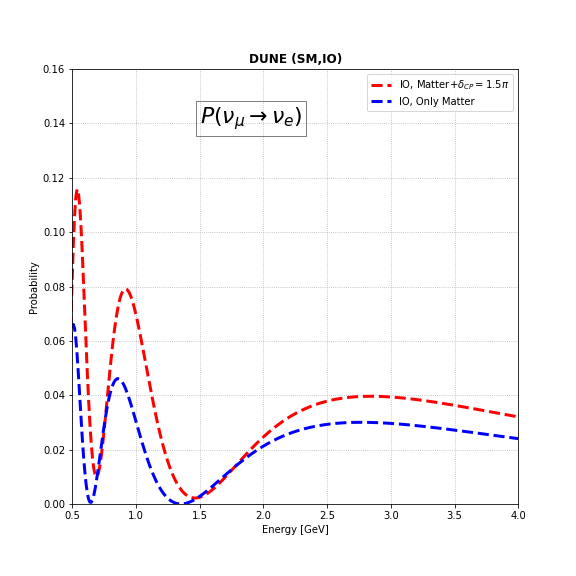}
\endminipage

\minipage{0.45\textwidth}
  \includegraphics[width=8.0cm,height=8.0cm]{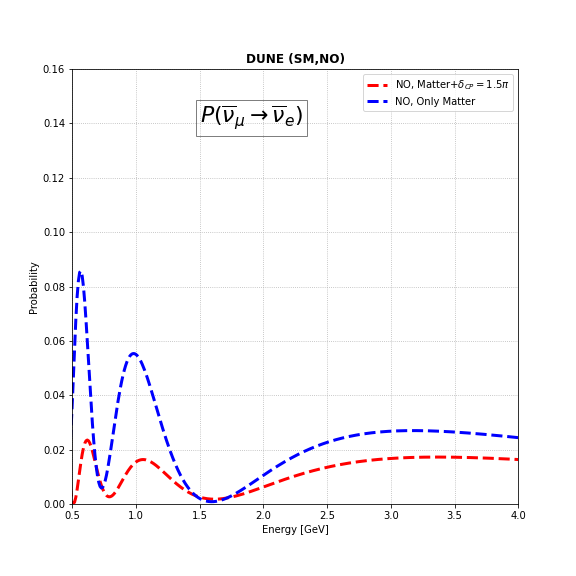}
\endminipage\hfill
\minipage{0.45\textwidth}
  \includegraphics[width=8.0cm,height=8.0cm]{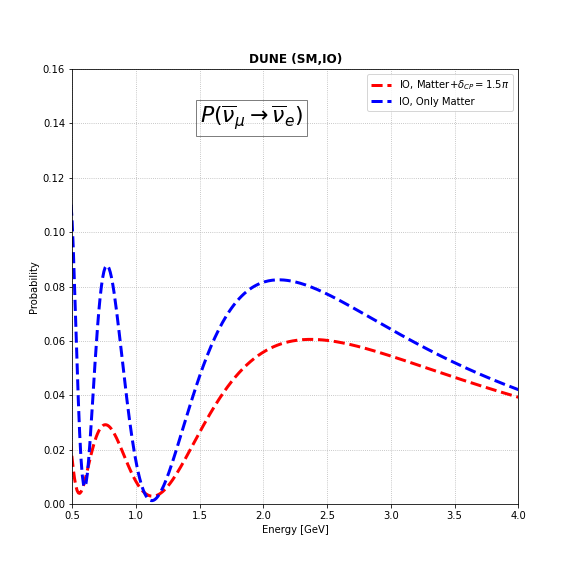}
\endminipage
\caption{Oscillation probability plots versus energy for NO (left) and IO (right) in the presence of matter for standard model parameter $\delta_{CP}=0$ and $\delta_{CP}=1.5\pi$ for neutrino (top panel) and anti-neutrino (bottom panel) sectors in the DUNE experimental setup.}
\end{figure}

In Figure 5 and Figure 6, we show the $A_{CP}$ parameter for energy varying from 0 to 4 GeV in normal mass (figure 5) and inverted mass (figure 6) hierarchy for the DUNE experiment. In these plots, we have plotted CP asymmetry in the presence of vacuum, matter, and non-standard interaction (NSI) arising from $\epsilon_{e\mu}$ and $\epsilon_{e\tau}$ sectors simultaneously. Clearly, at around DUNE energy, we can see the matter profile and dual NSI scenario probing opposite signs of CP asymmetry for NO and IO cases. Thus, hinting at the possibility of differentiating both the mass ordering in the DUNE experimental setup. The average $A_{CP}$ values for both the mass ordering in the presence of vacuum, matter, and dual NSI are depicted in Table II.

\vspace{10mm}

\begin{figure}[htbp]
  \includegraphics[width=14.0cm,height=10.5cm]{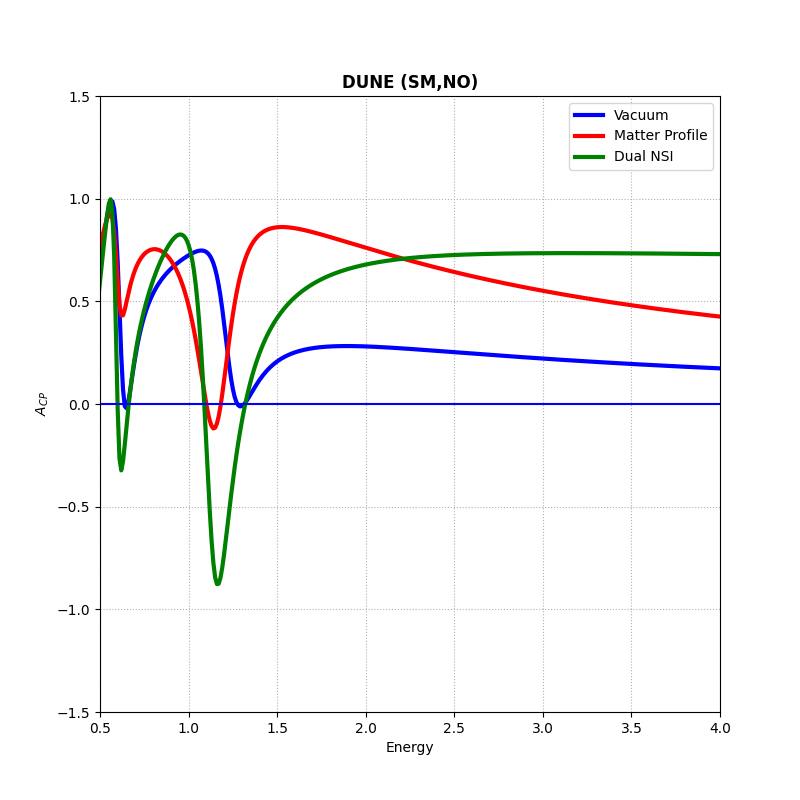}
\caption{CP asymmetry $A_{CP}$ versus Energy [in GeV] plot for NO scenario. In the above plots, we have included: vacuum, SM with matter effects, and SM with the inclusion of dual NSI arising from $\epsilon_{e\mu}$ and $\epsilon_{e\tau}$ simultaneously scenarios in the DUNE experimental setup.}
\end{figure}

\begin{figure}[htbp]
  \includegraphics[width=14.0cm,height=10.5cm]{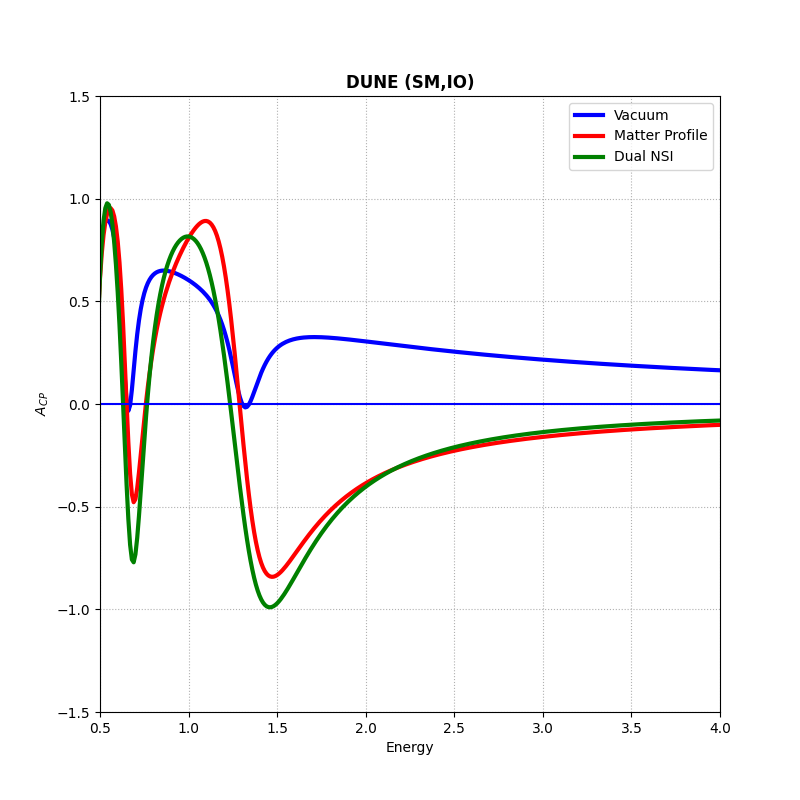}
\caption{CP asymmetry $A_{CP}$ versus Energy [in GeV] plot for IO scenario. In the above plots, we have included: vacuum, SM with matter effects, and SM with the inclusion of dual NSI arising from $\epsilon_{e\mu}$ and $\epsilon_{e\tau}$ simultaneously scenarios in the DUNE experimental setup.}
\end{figure}

\begin{table}[h!]
\caption{\label{tab:table2}{The average CP asymmetry (depicted in Figure 5 and Figure 6) in the presence of vacuum, matter, and dual NSI scenario for the DUNE energy window (2 GeV - 3 GeV) are included in the below table.}}
\begin{center}

\begin{tabular}{||c|cc||}
\hline
 &  \textbf{NO} &  \textbf{$A_{CP}$} ($\%$) \\ [1ex]  \hline
In & Vacuum & 25  \\
Presence & Matter  &  66 \\ 
of & Dual NSI  & 71 \\ [1ex]
\hline
 & \textbf{IO} & \textbf{$A_{CP}$}  ($\%$)  \\ [1ex]\hline
In & Vacuum & 26 \\
Presence & Matter  & -27\\
of & Dual NSI  & -27\\ 
\hline
\end{tabular}
\end{center}
\end{table}

\clearpage

\subsection{T2HK}

In this subsection of the analysis, we use T2HK running for 3 years and 4 years in $\nu$ mode and in $\Bar{\nu}$ mode, respectively. The T2HK experiment will use a 225 kt water Cherenkov detector. It will use an improved 30 GeV J-PARC beam with a power of 1.3 MW and a detector 295 km away from the source. The detector is located at 2.5$^{\circ}$ off-axis, and the neutrino flux peaks at around 0.6 GeV. 

In Figure 7, we show T2HK standard model oscillation probability plots versus energy for NO in vacuum for neutrino (top panel) and anti-neutrino (bottom panel) sectors. In the top right panel, we have neutrino oscillation probability versus energy plots for $\delta_{CP}=0$ and $\delta_{CP}=232^{\circ}$ (from nuFIT v5.2). On the left-hand side, we have the effective oscillation probability difference ($P_{\delta=232^{\circ}}$-$P_{\delta=0^{\circ}}$) plots.  In the case of the T2HK, we consider an energy window of 0.5 GeV to 1 GeV. We have restricted ourselves to the energy range around the peak neutrino beam for the sake of illustration. Here, the average positive probability difference in T2HK's energy window for neutrino is quite feeble (0.7$\%$). Similarly, in the case of anti-neutrino, the right side bottom plot gives the oscillation probability versus energy plot for $\delta_{CP}=0$ and $\delta_{CP}=232^{\circ}$. Here, the average negative probability difference in T2HK's energy window for anti-neutrino is also quite negligible (0.7$\%$). Here, the average probability difference value quantifies the change in probability with a change in the $\delta_{CP}$ phase in the presence of vacuum.

\begin{figure}[htbp]
\minipage{0.50\textwidth}
   \includegraphics[width=8.0cm,height=8.0cm]{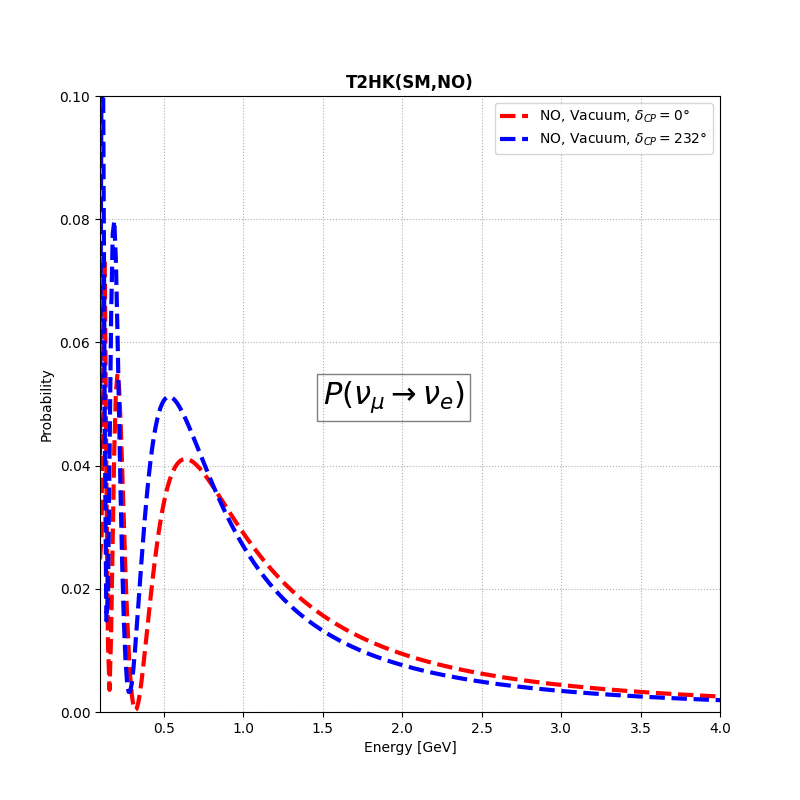}
\endminipage\hfill
\minipage{0.50\textwidth}
   \includegraphics[width=8.0cm,height=8.0cm]{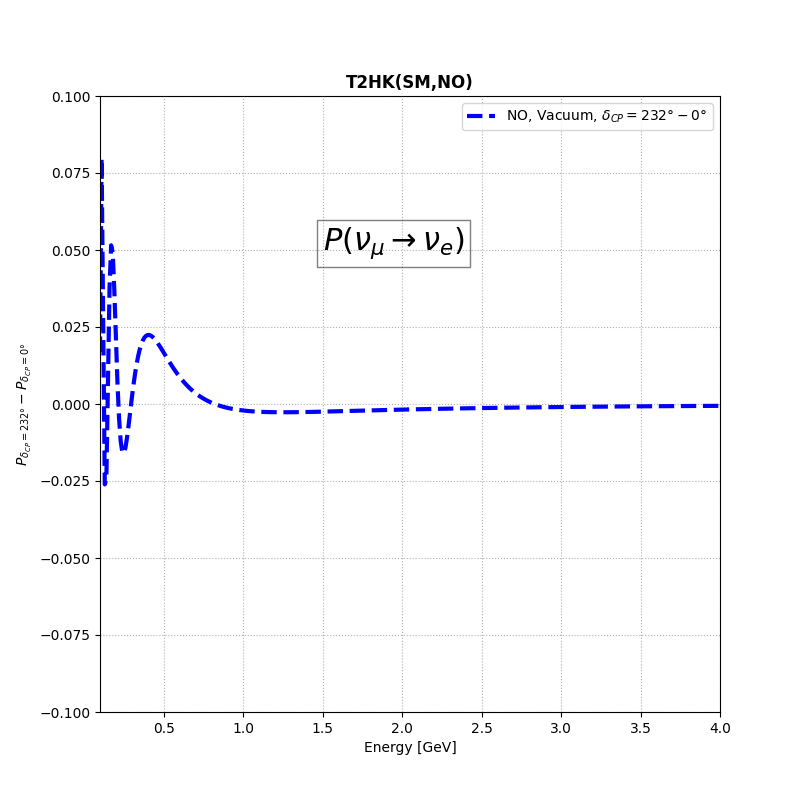}
\endminipage

\minipage{0.50\textwidth}
  \includegraphics[width=8.0cm,height=8.0cm]{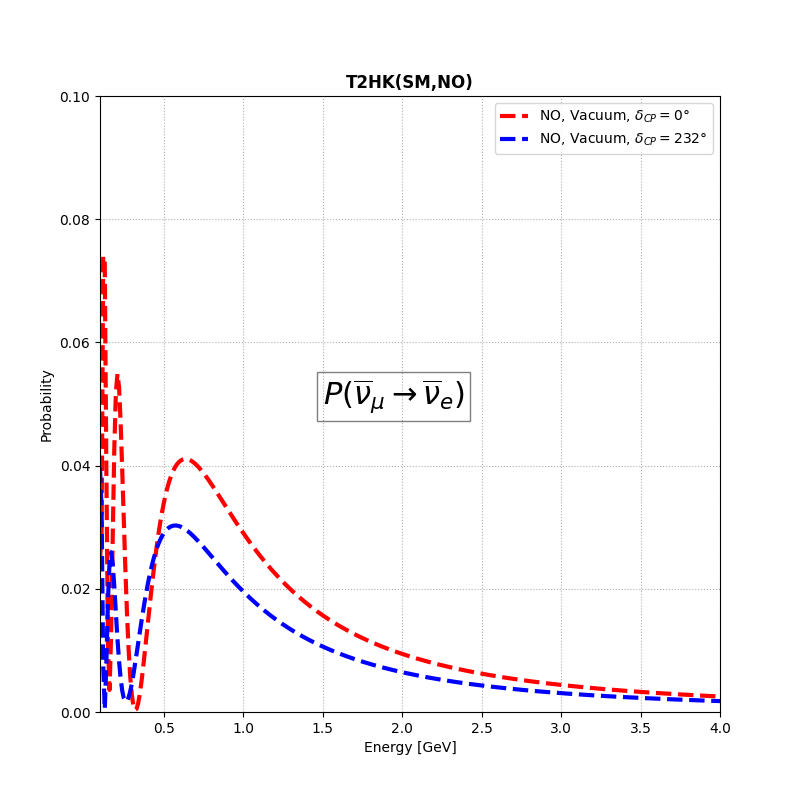}
\endminipage\hfill
\minipage{0.50\textwidth}
  \includegraphics[width=8.0cm,height=8.0cm]{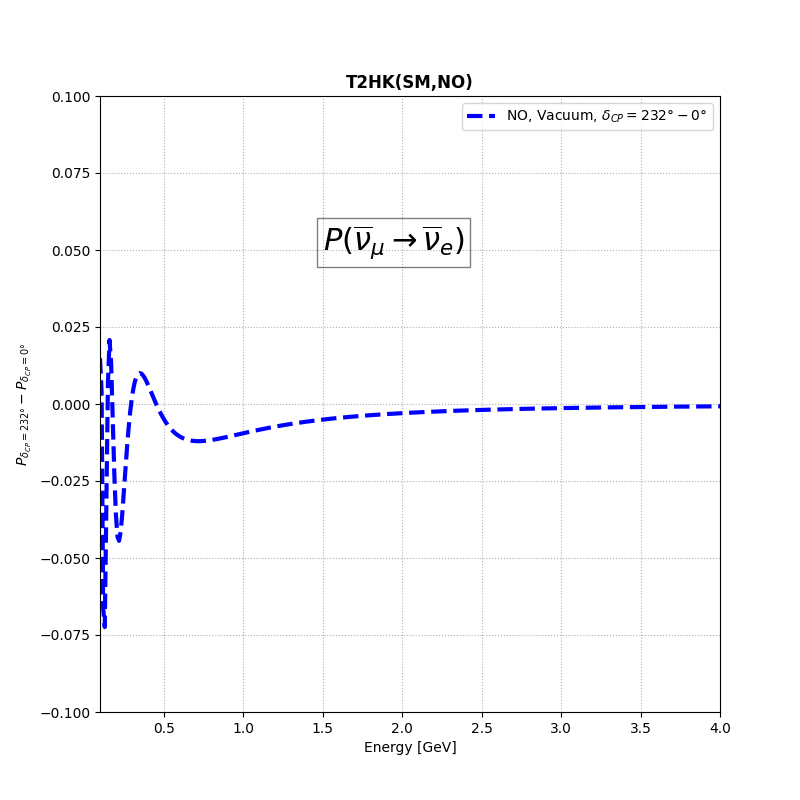}
\endminipage
\caption{Oscillation probability plots versus energy for NO in vacuum for neutrino (top panel) and anti-neutrino (bottom panel) sectors in T2HK experimental setup. Alongside oscillation  probability plots, we have oscillation probability difference plots ($P_{\delta=232^{\circ}}$-$P_{\delta=0^{\circ}}$) for neutrino(top right) and anti-neutrino (bottom right)}
\end{figure}

In Figure 8, we show now plotted oscillation probability plots in the presence of matter as well as vacuum for NO in the case of neutrino and anti-neutrino sectors in T2HK's experimental setup. The left side plots indicate oscillation probability versus energy ranging from 0 to 4 GeV, whereas the plots on the right depict the difference in oscillation probability between vacuum and matter. Using non-zero SM $\delta_{CP}=232^{\circ}$, as evident from the top left plot, the oscillation probability of the neutrino in the presence of matter is more dominating than in the case of vacuum at around T2HK's peak energy. The top right plot quantifies that the average difference between $P_{Vacuum}$ and $P_{Matter}$ for the neutrino sector is around -0.3$\%$. Similarly, in the case of anti-neutrino (bottom), the oscillation probability in the vacuum is more dominating than the presence of matter. Here, the average difference between $P_{Vacuum}$ and $P_{Matter}$ for the anti-neutrino sector is around 0.25$\%$. n these plots, we compare the oscillation probabilities in the case of T2HK when neutrinos travel through the vacuum and in the presence of matter medium.

\begin{figure}[htbp]
\minipage{0.50\textwidth}
   \includegraphics[width=8.0cm,height=8.0cm]{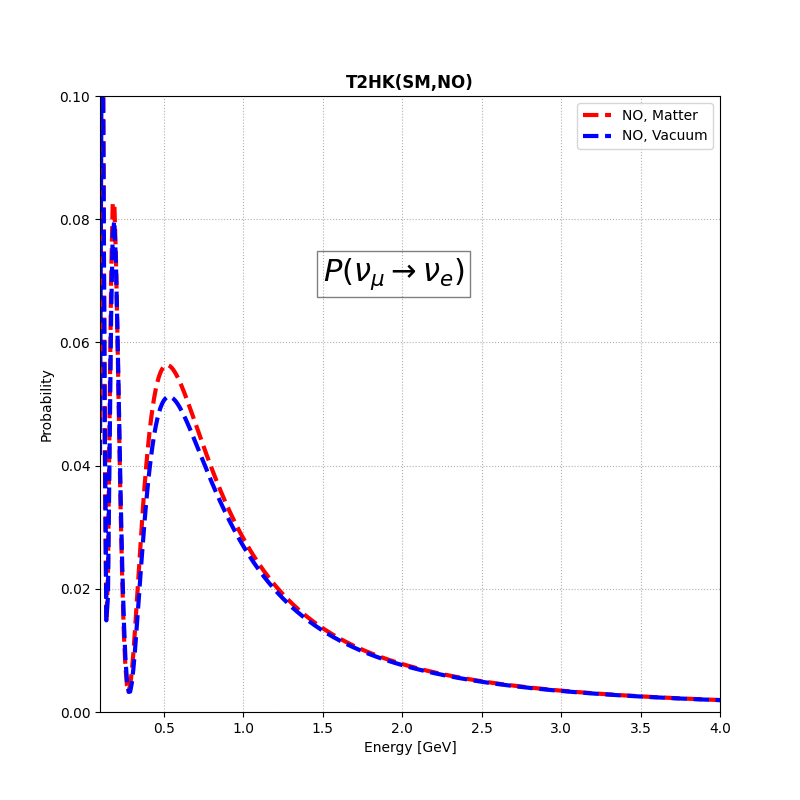}
\endminipage\hfill
\minipage{0.50\textwidth}
   \includegraphics[width=8.0cm,height=8.0cm]{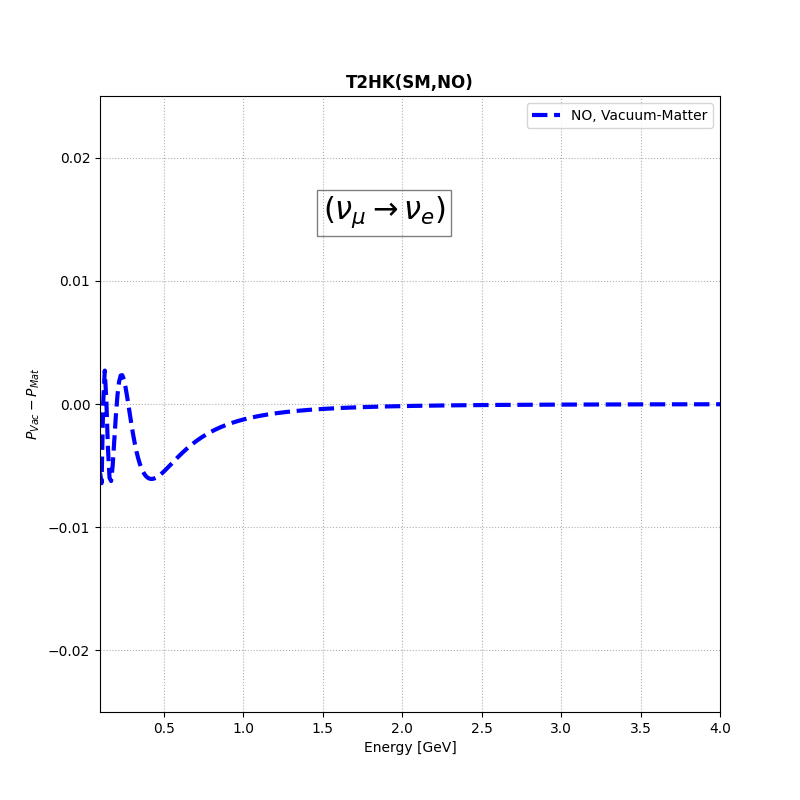}
\endminipage

\minipage{0.50\textwidth}
  \includegraphics[width=8.0cm,height=8.0cm]{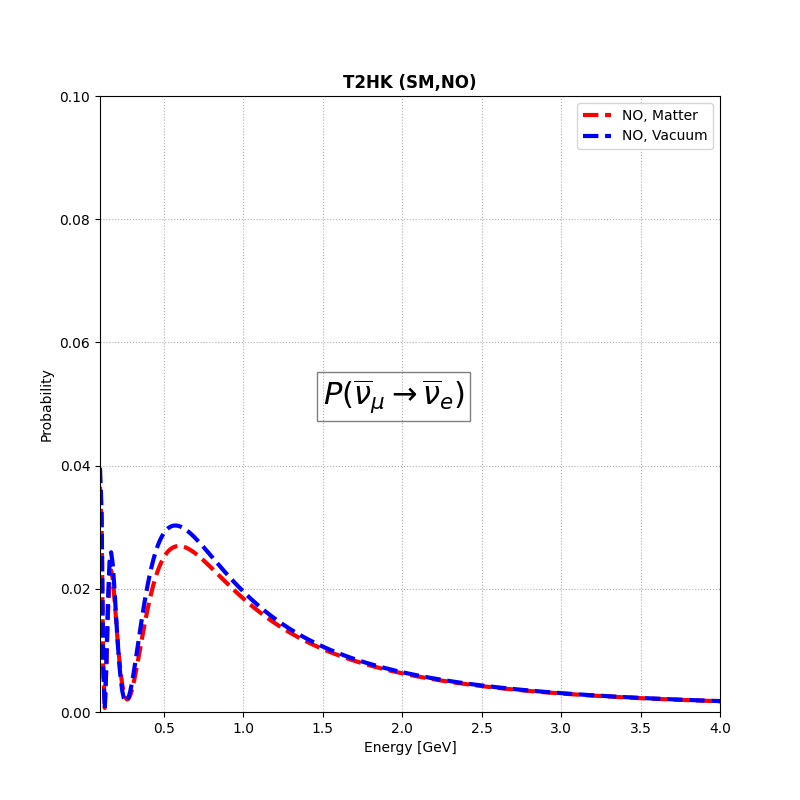}
\endminipage\hfill
\minipage{0.50\textwidth}
  \includegraphics[width=8.0cm,height=8.0cm]{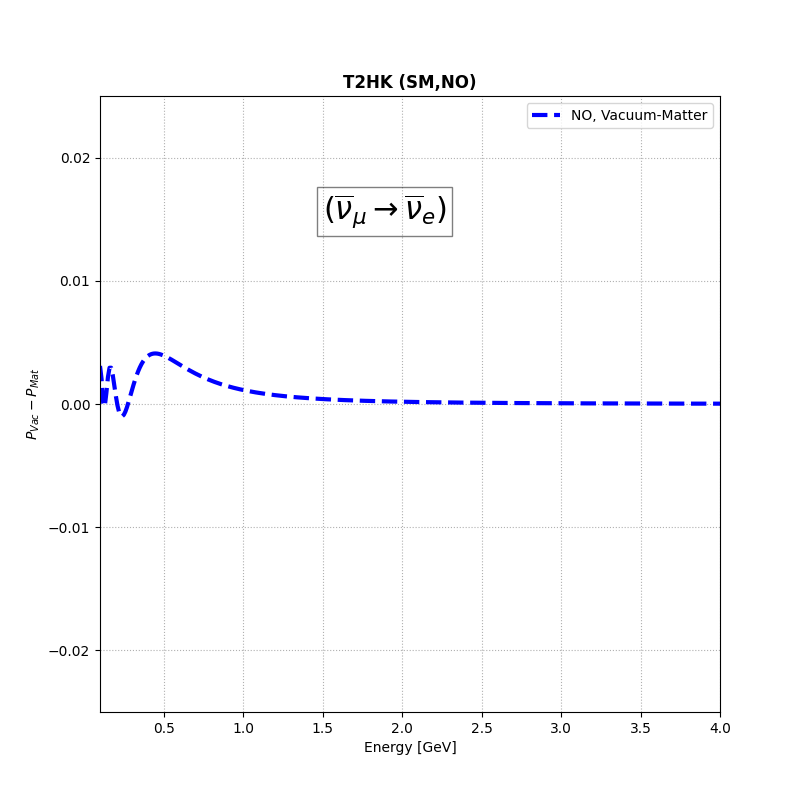}
\endminipage
\caption{Oscillation probability plots versus energy for NO in the vacuum and in the presence of matter for neutrino (top panel) and anti-neutrino (bottom panel) sectors in T2HK experimental setup. Alongside oscillation  probability plots, we have oscillation probability difference plots ($P_{Vacuum}$-$P_{Matter}$) for neutrino(top right) and anti-neutrino (bottom right)}
\end{figure}

In Figure 9, we show now plotted similar oscillation probability plots in the presence of matter as well as vacuum, but now for IO in the case of neutrino and anti-neutrino sectors in T2HK's experimental setup. The left side plots indicate oscillation probability versus energy ranging from 0 to 4 GeV, whereas the plots on the right depict the difference in oscillation probability between vacuum and matter. Using non-zero SM $\delta_{CP}=232^{\circ}$, as evident from the top left plot, the oscillation probability of the neutrino in the vacuum is more dominating than the oscillation probability in the presence of matter ($P_{Vacuum}$-$P_{Matter}=$ 0.42$\%$). Similarly, in the case of anti-neutrino(bottom), the oscillation probability in the matter is more dominating than the vacuum ($P_{Vacuum}$-$P_{Matter}=$ -0.33$\%$). Similar to the NO scenario, here, also we compare the oscillation probabilities when neutrinos travel through the vacuum and in the presence of matter medium.

\begin{figure}[htbp]
\minipage{0.50\textwidth}
   \includegraphics[width=8.0cm,height=8.0cm]{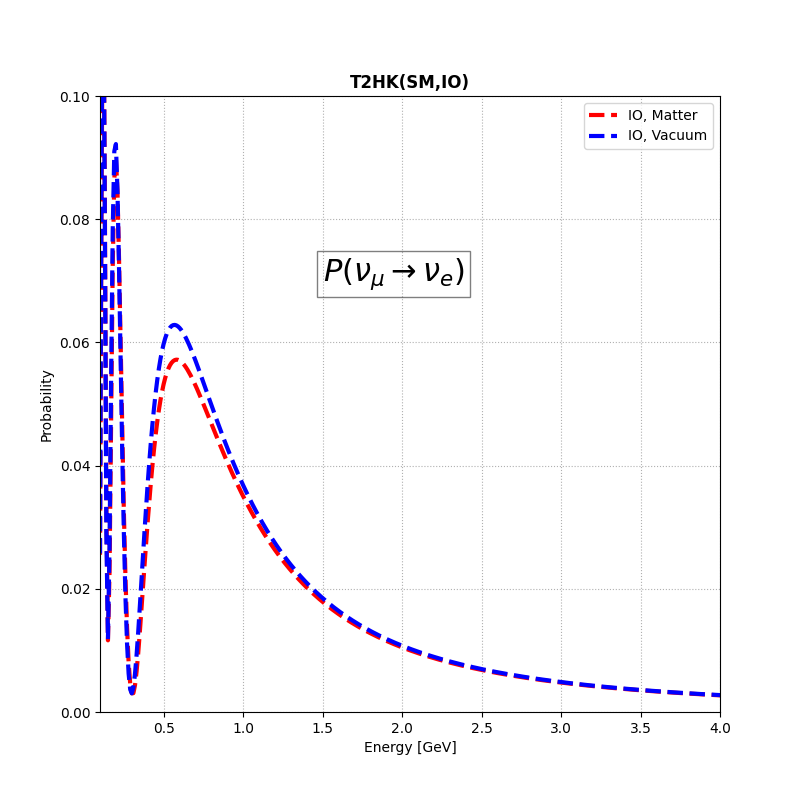}
\endminipage\hfill
\minipage{0.50\textwidth}
   \includegraphics[width=8.0cm,height=8.0cm]{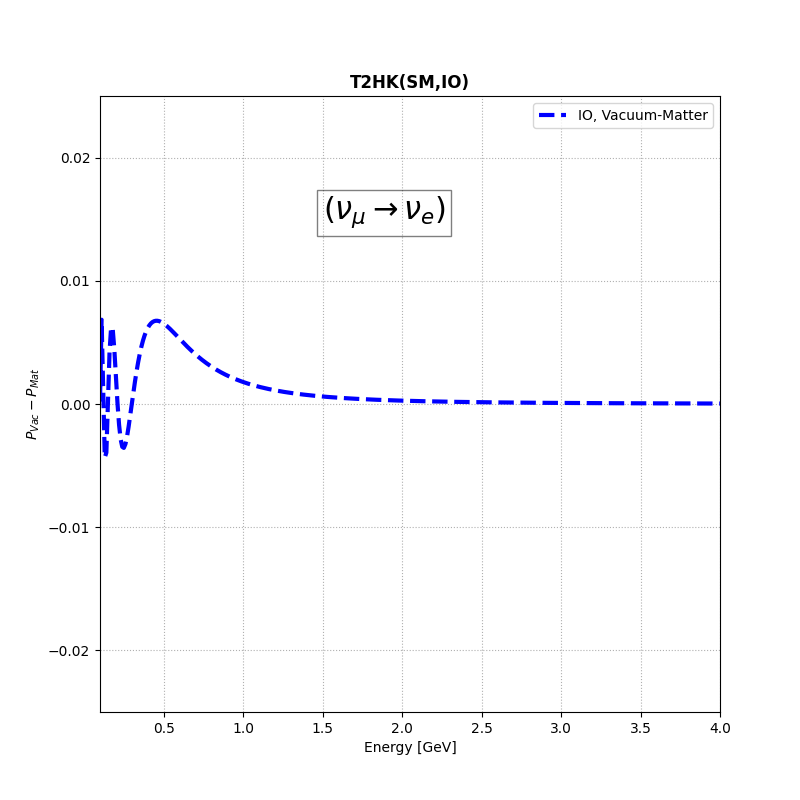}
\endminipage

\minipage{0.50\textwidth}
  \includegraphics[width=8.0cm,height=8.0cm]{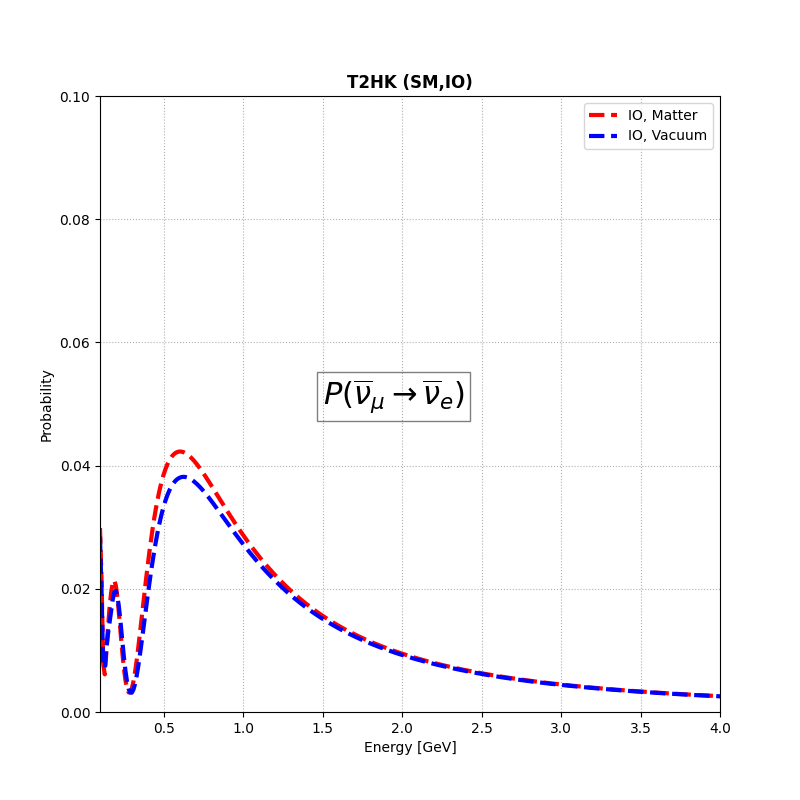}
\endminipage\hfill
\minipage{0.50\textwidth}
  \includegraphics[width=8.0cm,height=8.0cm]{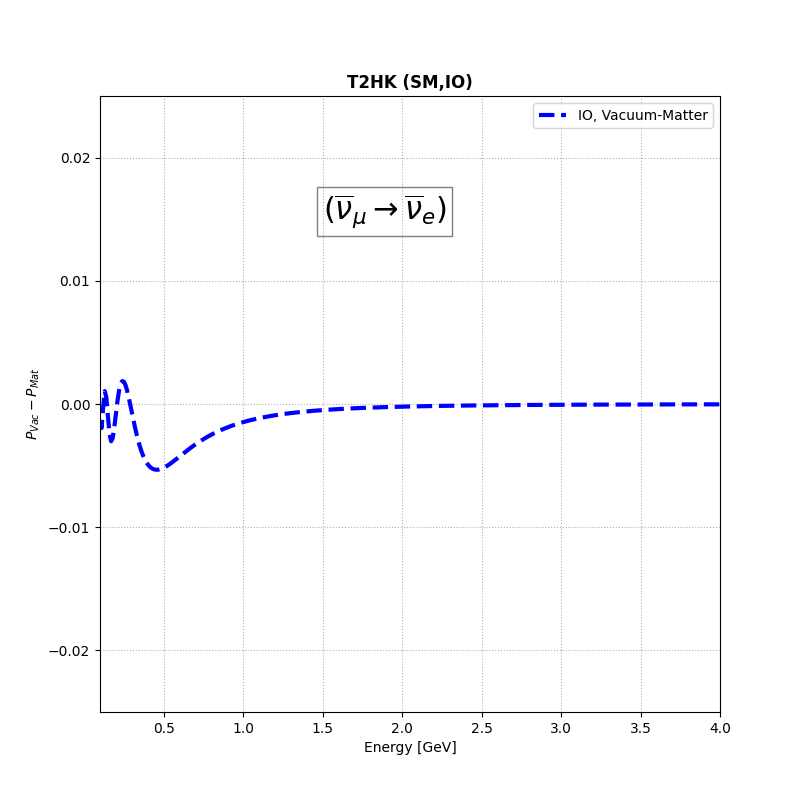}
\endminipage
\caption{Oscillation probability plots versus energy for IO in the vacuum and in the presence of matter for neutrino (top panel) and anti-neutrino (bottom panel) sectors in T2HK experimental setup. Alongside oscillation  probability plots, we have oscillation probability difference plots ($P_{Vacuum}$-$P_{Matter}$) for neutrino (top right) and anti-neutrino (bottom right)}
\end{figure}

Similar to DUNE, in the case of T2HK also, we have considered the SM $\delta_{CP}$ to be 232$^{\circ}$ for NO and 276$^{\circ}$ for IO in all the previous and subsequent plots. In this particular figure, we have considered a nominal $\delta_{CP}$ value of 1.5$\pi$ for NO and IO. Thus, in Figure 10, we utilise the non-zero $\delta_{CP}$ value as given by T2K, i.e., 1.5$\pi$.In this figure, we show oscillation probability versus energy plots for neutrino NO (top left) as well as IO (top right) and also for anti-neutrino NO (bottom left) and IO (bottom right). In the neutrino NO case, the plot indicates that the oscillation probability in the presence of matter effect along with non-zero standard model $\delta_{CP}$ phase of 1.5$\pi$ is more dominating than the oscillation probabilities in the presence of matter effect with zero $\delta_{CP}$ phase at around T2HK's energy window energy. A similar dominating feature is also present in the neutrino IO case. For the anti-neutrino case, in the NO scenario, the oscillation probabilities in the presence of matter effect with zero $\delta_{CP}$ phase is more dominating than in the presence of matter effect with non-zero $\delta_{CP}$ phase of 1.5$\pi$. Again a similar feature is visualised in the plot for the anti-neutrino IO case. All the above cases are illustrated at T2HK's energy window of 0.5 GeV - 1 GeV. Here, we get the picture of a change in the oscillation probabilities with nominal 1.5$\pi$ and zero $\delta_{CP}$ phase in the presence of matter.

\begin{figure}[htbp]
\minipage{0.45\textwidth}
  \includegraphics[width=8.0cm,height=8.0cm]{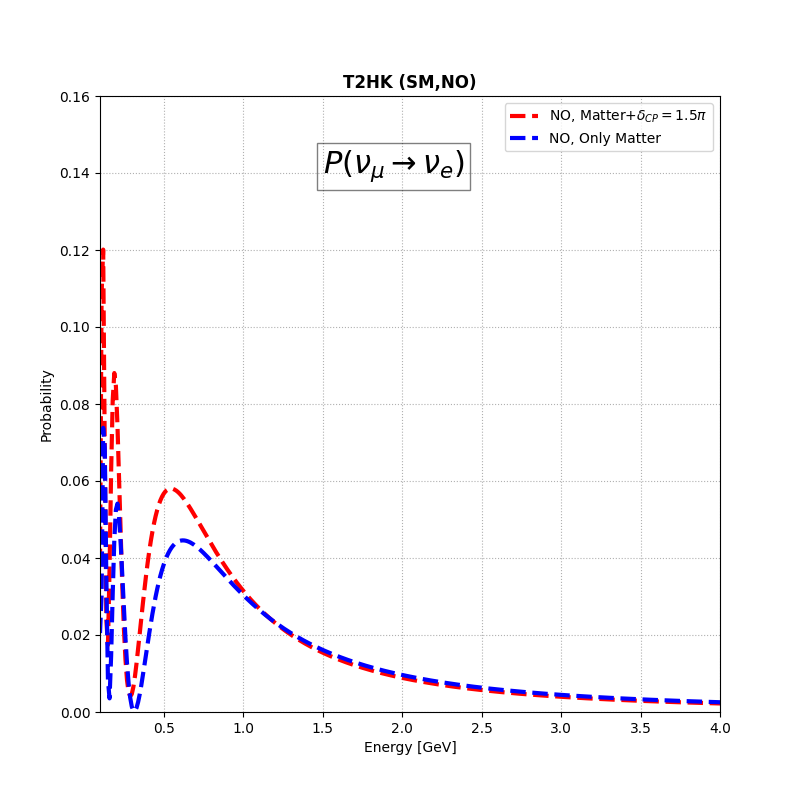}
\endminipage\hfill
\minipage{0.45\textwidth}
  \includegraphics[width=8.0cm,height=8.0cm]{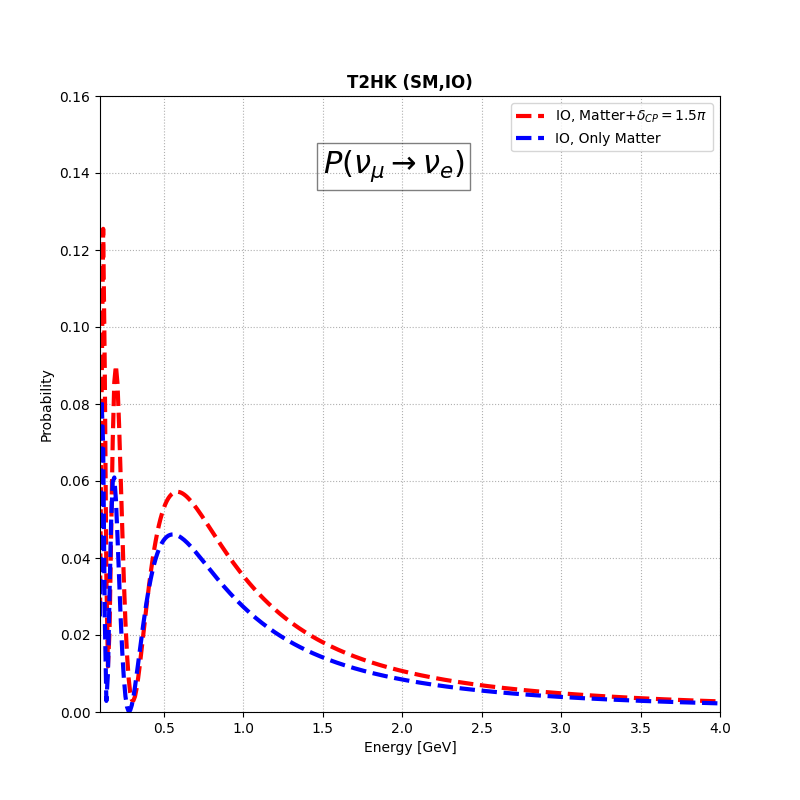}
\endminipage

\minipage{0.45\textwidth}
  \includegraphics[width=8.0cm,height=8.0cm]{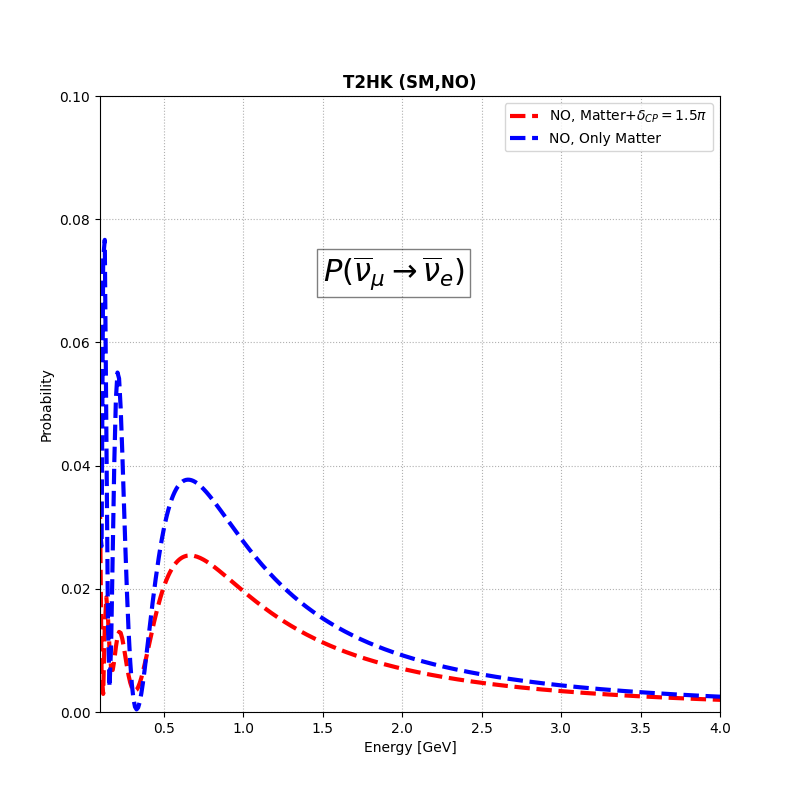}
\endminipage\hfill
\minipage{0.45\textwidth}
\includegraphics[width=8.0cm,height=8.0cm]{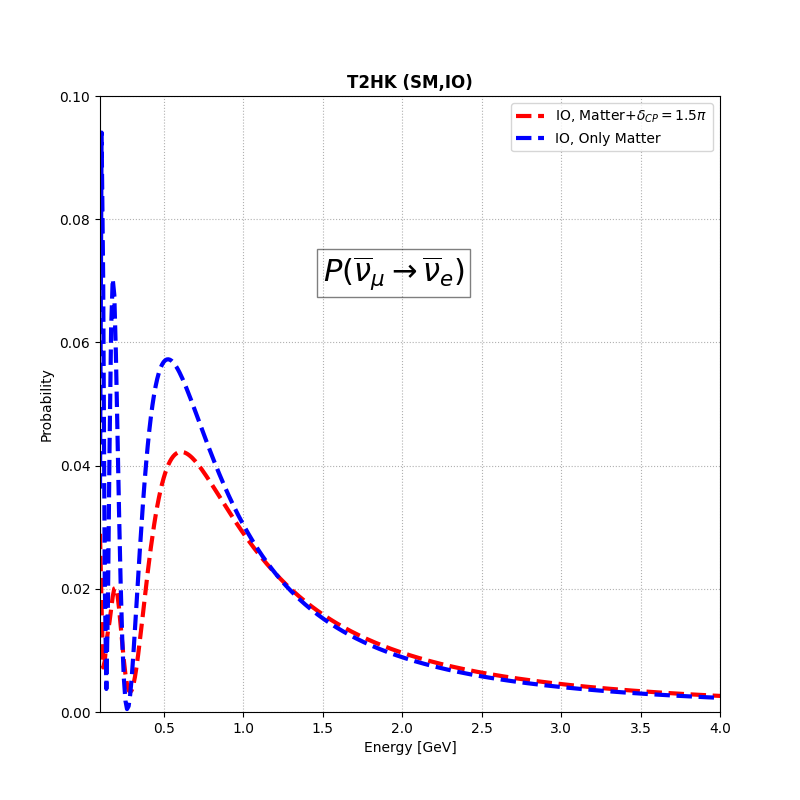}
\endminipage
\caption{Oscillation probability plots versus energy for NO (left) and IO (right) in the presence of matter for standard model parameter $\delta_{CP}=0$ and $\delta_{CP}=1.5\pi$ for neutrino (top panel) and anti-neutrino (bottom panel) sectors in the T2HK experimental setup.}
\end{figure}

In Figure 11 and Figure 12,  we exhibit the $A_{CP}$ parameter for energy varying from 0 to 4 GeV in normal mass (figure 11) and inverted mass (figure 12) hierarchy for the T2HK experimental setup. In these plots, we have plotted CP asymmetry in the presence of vacuum, matter, and non-standard interaction (NSI) arising from $\epsilon_{e\mu}$ and $\epsilon_{e\tau}$ sectors simultaneously. At around the T2HK's energy window, we can see the matter profile and dual NSI scenario probing more or less similar signs of CP asymmetry for both NO and IO cases. Thus, hinting at the possibility that the T2HK experiment might not be able to differentiate between the mass orderings. The average $A_{CP}$ values for both the mass ordering in the presence of vacuum, matter, and dual NSI are depicted in Table III.

\begin{figure}[hbt!]
  \includegraphics[width=14.0cm,height=12.0cm]{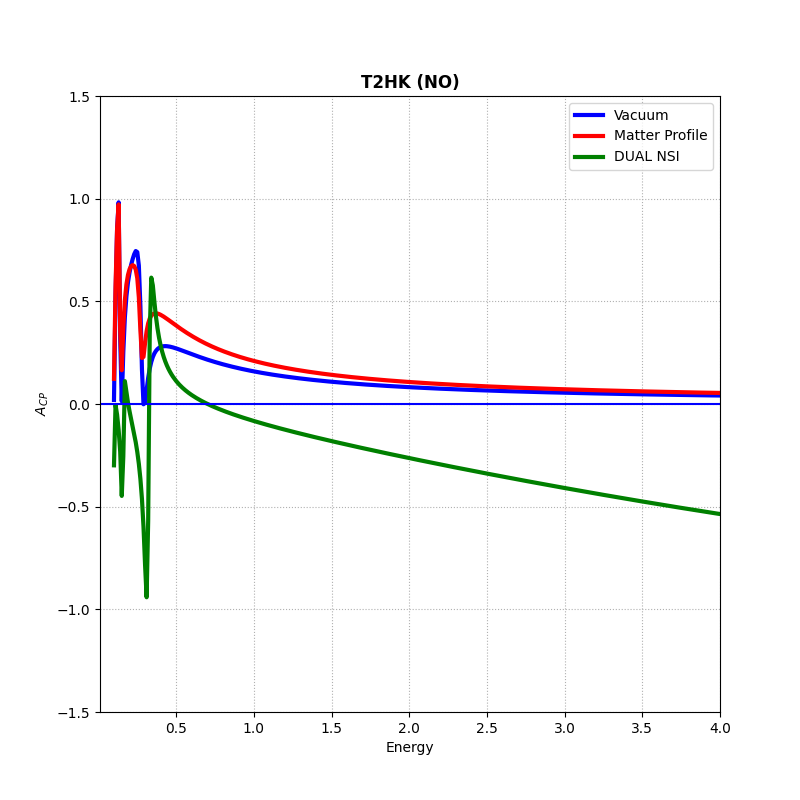}
\caption{CP asymmetry $A_{CP}$ versus Energy [in GeV] plot for NO scenario. In the above plots, we have included: vacuum, SM with matter effects, and SM with the inclusion of dual NSI arising from $\epsilon_{e\mu}$ and $\epsilon_{e\tau}$ simultaneously scenarios in the T2HK experimental setup.}
\end{figure}

\begin{figure}[hbt!]
  \includegraphics[width=14.0cm,height=12.0cm]{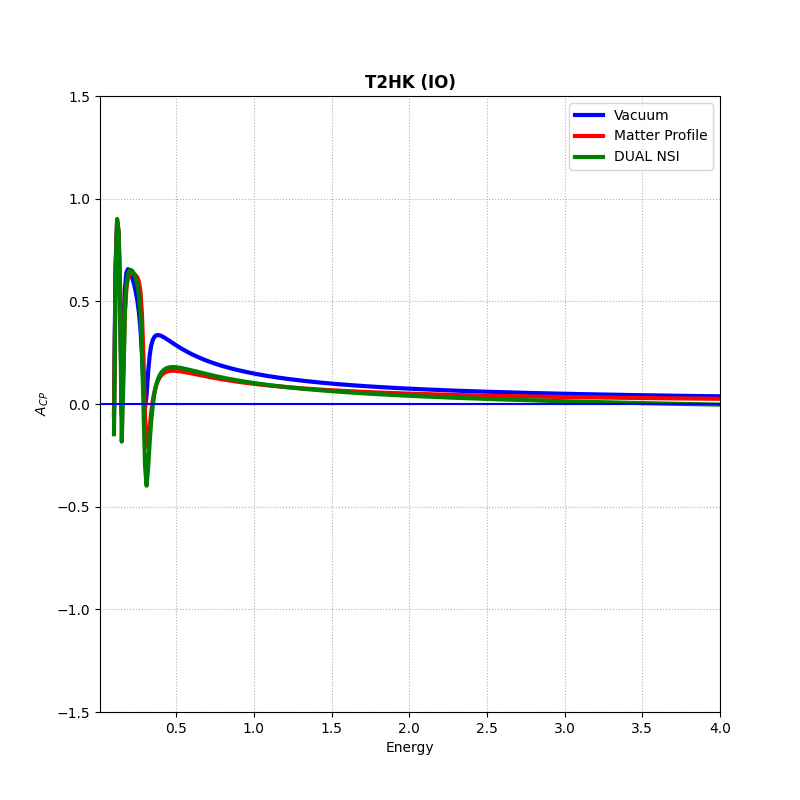}
\caption{CP asymmetry $A_{CP}$ versus Energy [in GeV] plot for IO scenario. In the above plots, we have included: vacuum, SM with matter effects, and SM with the inclusion of dual NSI arising from $\epsilon_{e\mu}$ and $\epsilon_{e\tau}$ simultaneously scenarios in the T2HK experimental setup.}
\end{figure}

\begin{table}[h!]
\caption{\label{tab:table3}{The average CP asymmetry (depicted in Figure 11 and Figure 12) in the presence of vacuum, matter, and dual NSI scenario for the T2HK energy window (0.5 GeV - 1 GeV) are included in the below table.}}
\begin{center}

\begin{tabular}{||c|cc||}
\hline
 &  \textbf{NO} &  \textbf{$A_{CP}$} ($\%$) \\ [1ex]  \hline
In & Vacuum & 21  \\
Presence & Matter  &  30 \\ 
of & Dual NSI  & 2 \\ [1ex]
\hline
 & \textbf{IO} & \textbf{$A_{CP}$}  ($\%$)  \\ [1ex]\hline
In & Vacuum & 22  \\
Presence & Matter  & 13\\
of & Dual NSI  & 14\\ 
\hline
\end{tabular}
\end{center}
\end{table}
\subsection{DUNE and T2HK}

 In Figure 13, the top panel shows CP asymmetry versus energy in the presence of SM with matter effects, and the bottom panel indicates a similar CP asymmetry but now with the inclusion of NSI from $\epsilon_{e\mu}$ and $\epsilon_{e\tau}$ sector simultaneously (dual NSI), for the DUNE (left) and T2HK (right) experimental setup. In the case of the DUNE, we consider an energy window of 2 GeV to 3 GeV, and for T2HK, 0.5 GeV to 1.0 GeV. We have restricted ourselves to the energy range around the peak neutrino beam for the sake of illustration. Henceforth, all the mentioned $A_{CP}$ parameter values are average values calculated in the above-mentioned energy window.
 
 In the case of DUNE, in the top panel, we found that the inverted mass ordering (IO) scenario prefers a  negative $A_{CP}$ value of 27$\%$, and the normal mass ordering (NO) shows a positive $A_{CP}$ value of 66$\%$. We consider here $\delta_{CP} = 276^{\circ}$ for the IO scenario and  $\delta_{CP} = 232^{\circ}$ for the NO case [taken from nuFIT v5.2]. In contrast, in the T2HK energy window, both normal mass ordering and inverted mass ordering show positive $A_{CP}$ values of 30$\%$ and 13$\%$, respectively. Now, with the inclusion of the dual NSI effect, and in the bottom panel, for the DUNE energy window, NO prefers a positive $A_{CP}$ value of 71$\%$, whereas IO prefers a negative $A_{CP}$ value of 27$\%$. Similar to the SM with matter effect case, T2HK in dual NSI scenario also prefers positive $A_{CP}$ values of 33$\%$ and 14$\%$ for NO and IO, respectively. It can be clearly seen that irrespective of the SM with matter effect or dual NSI case we consider, NO and IO always prefer the opposite $A_{CP}$ sign in the DUNE energy window and the same $A_{CP}$ sign in the T2HK energy window. Thus, in the presence of dual NSI in DUNE's experimental setup, we can clearly differentiate between the mass hierarchies if we measure the $A_{CP}$ value, which is not observed in the T2HK case.

\begin{figure}[htbp]
\minipage{0.45\textwidth}
  \includegraphics[width=8.0cm,height=8.0cm]{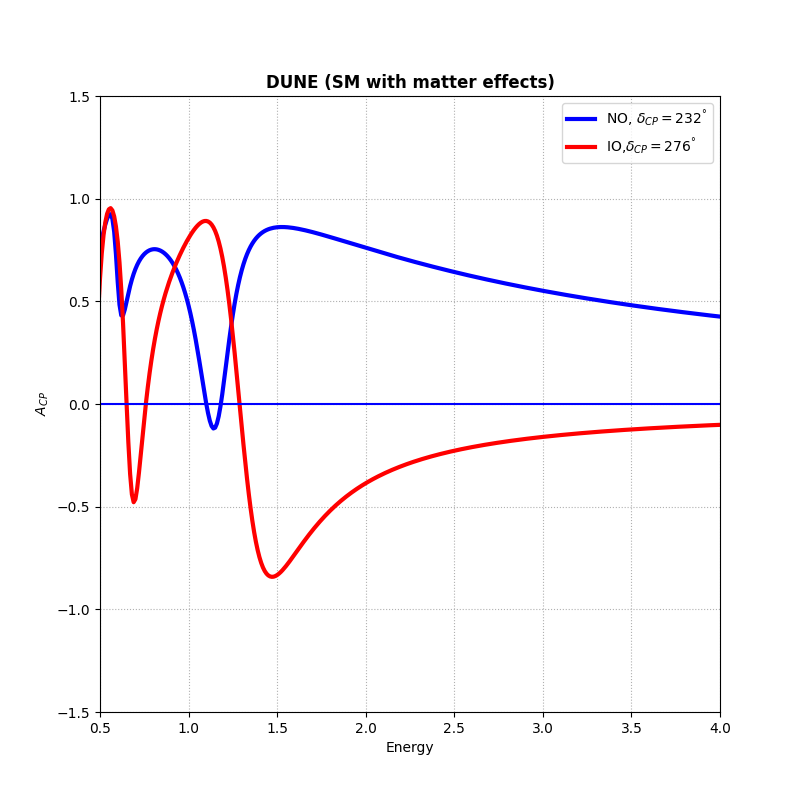}
\endminipage\hfill
\minipage{0.45\textwidth}
  \includegraphics[width=8.0cm,height=8.0cm]{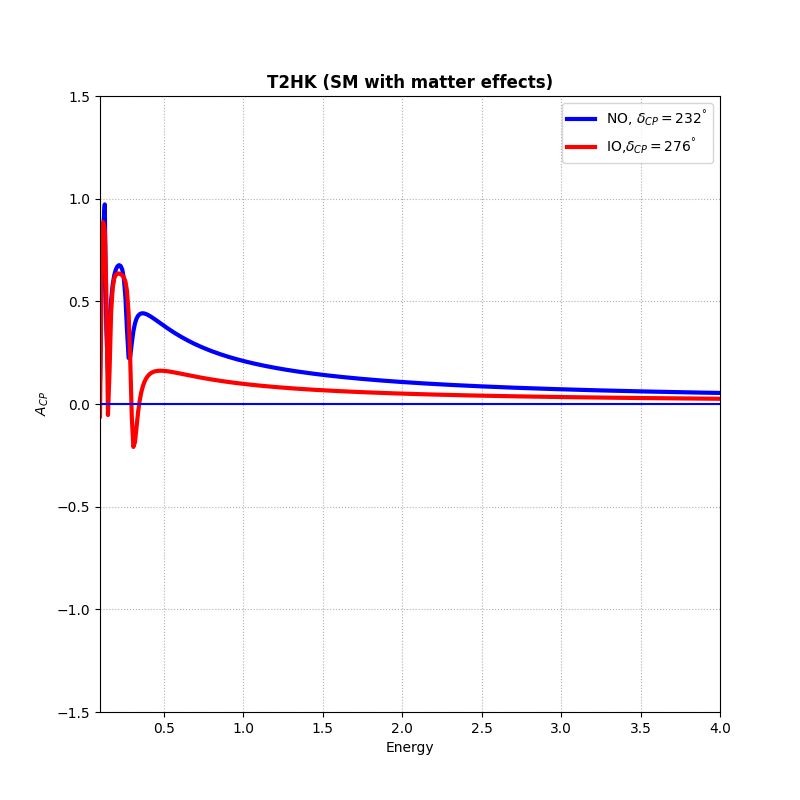}
\endminipage

\minipage{0.45\textwidth}
  \includegraphics[width=8.0cm,height=8.0cm]
  {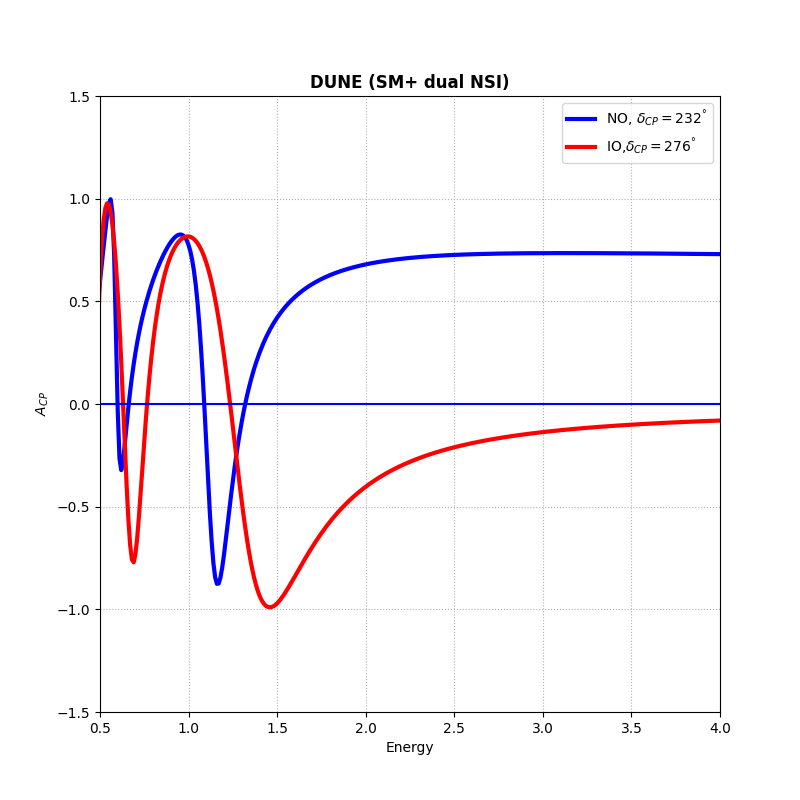}
\endminipage\hfill
\minipage{0.45\textwidth}
  \includegraphics[width=8.0cm,height=8.0cm]{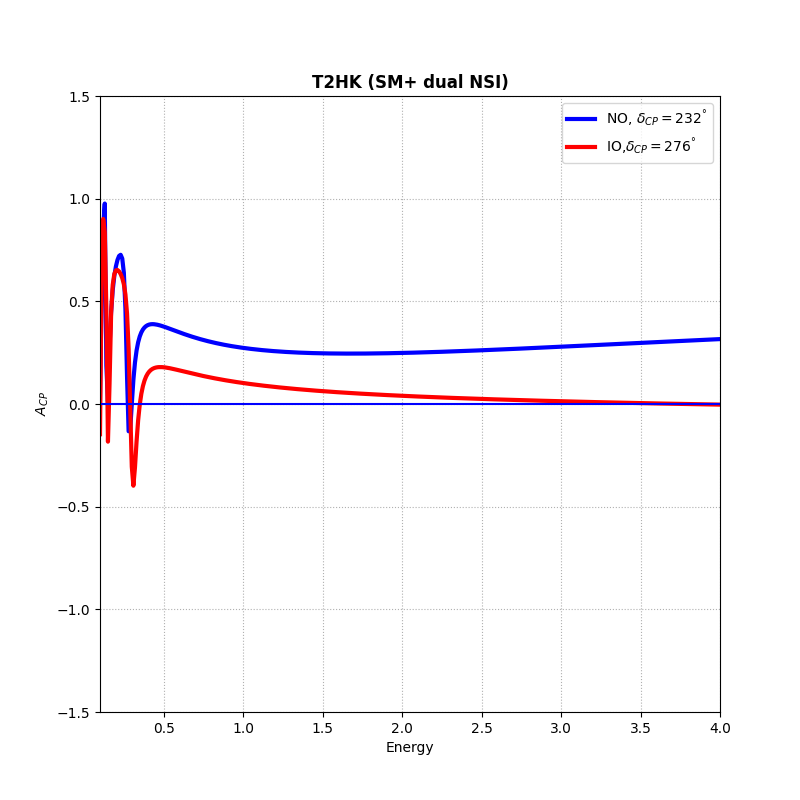}
\endminipage
\caption{CP asymmetry $A_{CP}$ versus energy [in GeV] in the presence of SM with matter effect (top) and dual NSI arising from the $\epsilon_{e\mu}$ and $\epsilon_{e\tau}$ sector simultaneously (bottom) in case of DUNE (left) and T2HK (right) experimental setup}
\end{figure}

In Figure 14, we show $A_{CP}$ versus SM parameter $\delta_{CP}$ varying from 0 to 2$\pi$. On the left side, one can see SM as well as SM along with dual NSI arising from the $\epsilon_{e\mu}$ and $\epsilon_{e\tau}$ sector simultaneously, plots for both NO and IO in the case of DUNE, whereas on the right side, along with SM, we have included dual NSI but now for T2HK. 

Here, for illustration, we consider the range of SM parameter $\delta_{CP}$ to be ranging from 1.2$\pi$ to 1.4$\pi$ for NO and 1.4$\pi$ to 1.6$\pi$ for IO, which is around their best-fit values from nuFIT v5.2. Hereafter, all the mentioned values of $A_{CP}$ are also the average values calculated in the above-mentioned $\delta_{CP}$ range for both the ordering scenarios. In the case of DUNE, the $A_{CP}$ parameter indicates a positive value of 62$\%$ for NO and a negative value of 22$\%$ for IO in the SM case. With the inclusion of dual NSI, the $A_{CP}$ parameter value indicates a positive value of 72$\%$ for NO and a negative value of 23$\%$ for IO. For T2HK, the $A_{CP}$ parameter indicates a positive value of 33$\%$ for NO and 14$\%$ for IO in the SM case. With the inclusion of dual NSI, the $A_{CP}$ parameter value now gives a positive value of 34$\%$ for NO and 15$\%$ for IO.
It is evident that the sign of $A_{CP}$ parameter will indicate the preferable mass ordering in nature, which looks to be possible in the upcoming DUNE experimental setup. T2HK does not show this interesting feature for mass ordering with the study of the $A_{CP}$ parameter. Dune, with its broader energy spectrum, larger base length, and more matter effect, will offer a better chance of visualising the difference in the mass hierarchies than T2HK.

\begin{figure}[htbp]
\minipage{0.45\textwidth}
  \includegraphics[width=9.0cm,height=9.0cm]{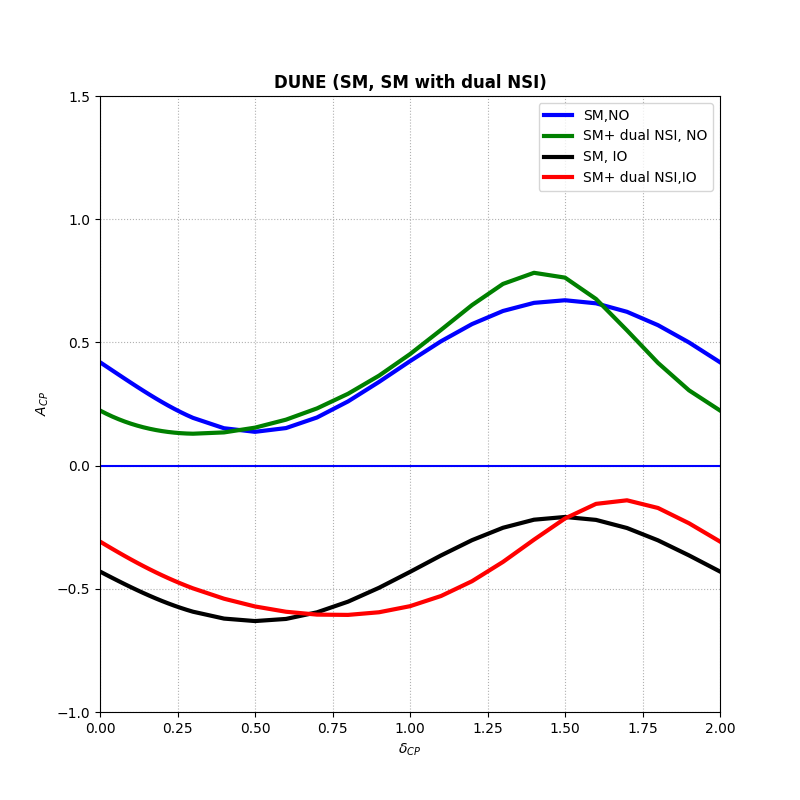}
\endminipage\hfill
\minipage{0.45\textwidth}
  \includegraphics[width=9.0cm,height=9.0cm]{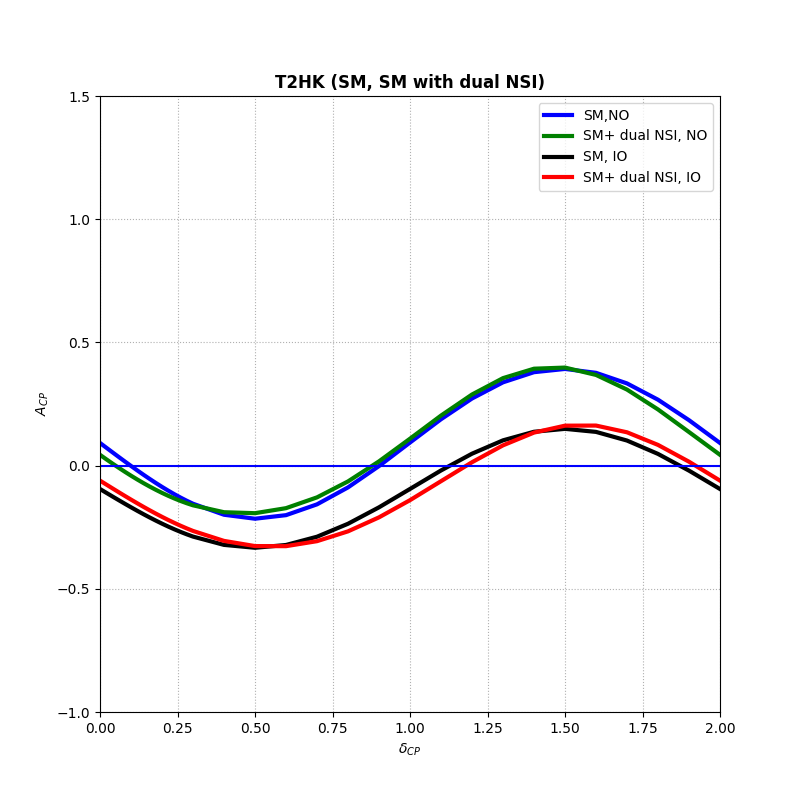}
\endminipage
\caption{CP asymmetry $A_{CP}$ versus standard model parameter $\delta_{CP}$ ranging from 0 to 2$\pi$ in the case of SM as well as in the presence of dual NSI arising from the $\epsilon_{e\mu}$ and $\epsilon_{e\tau}$ sector simultaneously, for DUNE (left) and T2HK (right) experimental setup.}
\end{figure}

In this analysis, we also study the $\Delta A_{\alpha \beta}(\delta_{CP}) \equiv A_{\alpha \beta}(\delta_{CP}\neq0) - A_{\alpha \beta}(\delta_{CP}=0)$ to understand the CP-violating effects in long-baseline neutrino oscillation experiments like DUNE and T2HK. The idea is to explore  whether we can also notice any such interesting signatures, as evident in the case of $A_{CP}$. As before, here we discuss $\Delta A_{\alpha \beta}^{CP}$ parameters for DUNE (left) and for T2HK (right) experiments. For normal mass ordering, we consider the non-zero $\delta_{CP}$ value to be 232$^{\circ}$ and for inverted mass hierarchy 276$^{\circ}$ both are taken from nuFIT v5.2.

In Figure 15, we exhibit $\Delta A_{\mu e}(\delta)$ observable with energy varying from 0 to 4 GeV. As discussed earlier, for DUNE, we consider an energy window of 2 GeV - 3 GeV, and for T2HK, 0.5 GeV to 1.0 GeV. Here also, all the mentioned $\Delta A_{\mu e}^{CP}(\delta)$ parameter values are average values calculated in the above-mentioned energy window.
The top panel is plotted in the presence of SM with matter effects, and the bottom panel includes dual NSI arising from $\epsilon_{e\mu}$ and $\epsilon_{e\tau}$ sectors simultaneously. As seen from the figure, in the case of SM with matter effects for DUNE (top left), NO indicates a  positive $\Delta A_{\mu e}(\delta)$ value of 16$\%$, whereas IO prefers a  positive $\Delta A_{\mu e}(\delta)$ value of 20$\%$ at around DUNE's energy window. With the inclusion of dual NSI, DUNE gives a positive $\Delta A_{\mu e}(\delta)$ value for both NO and IO. For the NO case, the preferred positive  $\Delta A_{\mu e}(\delta)$ value is 38$\%$, and for IO, it is 12$\%$. With energy beyond 3 GeV, we can see a clear distinction between normal mass and inverted mass hierarchy in DUNE. 

In the T2HK energy window, for SM with matter effect inclusion, both NO and IO prefer similar signs of $\Delta A_{\mu e}(\delta)$. More explicitly, for NO, the  positive $\Delta A_{\mu e}(\delta)$ value is about 21$\%$, and for IO, it is 22$\%$. Also, in the dual NSI case (bottom right figure), both NO and IO prefer a positive $\Delta A_{\mu e}(\delta)$ value, i.e., for NO,  $\Delta A_{\mu e}(\delta) = 31 \%$ and for IO,  $\Delta A_{\mu e}(\delta) = 16 \%$. In T2HK experimental setup, beyond 1 GeV energy, we can see a clear distinction between normal mass and inverted mass hierarchies as both the scenarios indicate opposite $\Delta A_{\mu e}(\delta)$ sign. 

It can be clearly seen that $A_{CP}$ observable plays a more important role than  $\Delta A_{\mu e}(\delta)$ to distinguish between both mass orderings at DUNE and T2HK's energy windows.

\begin{figure}[htbp]
\minipage{0.45\textwidth}
  \includegraphics[width=8.0cm,height=8.0cm]{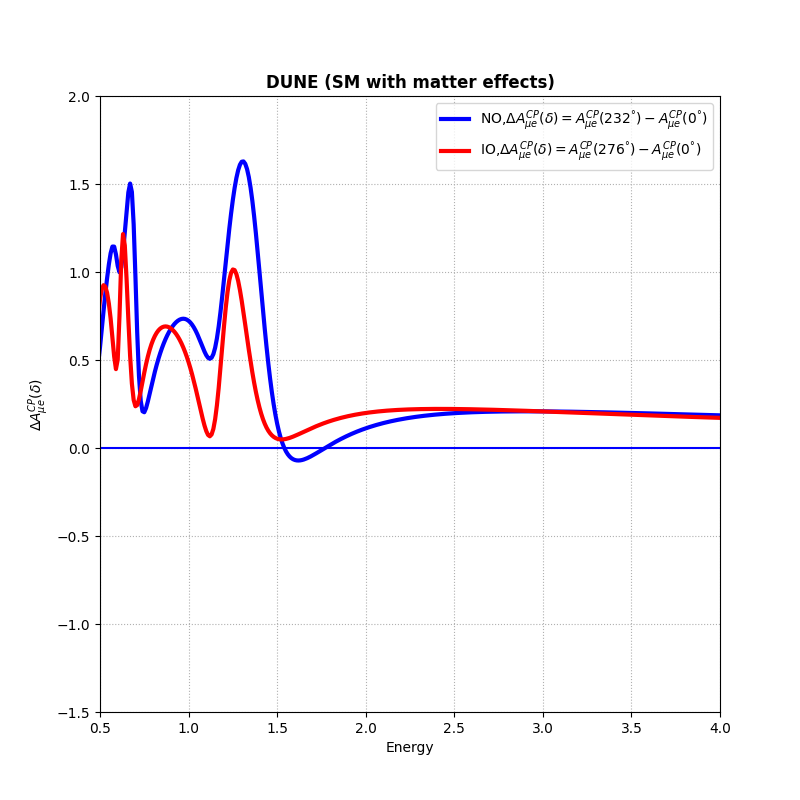}
\endminipage\hfill
\minipage{0.45\textwidth}
  \includegraphics[width=8.0cm,height=8.0cm]{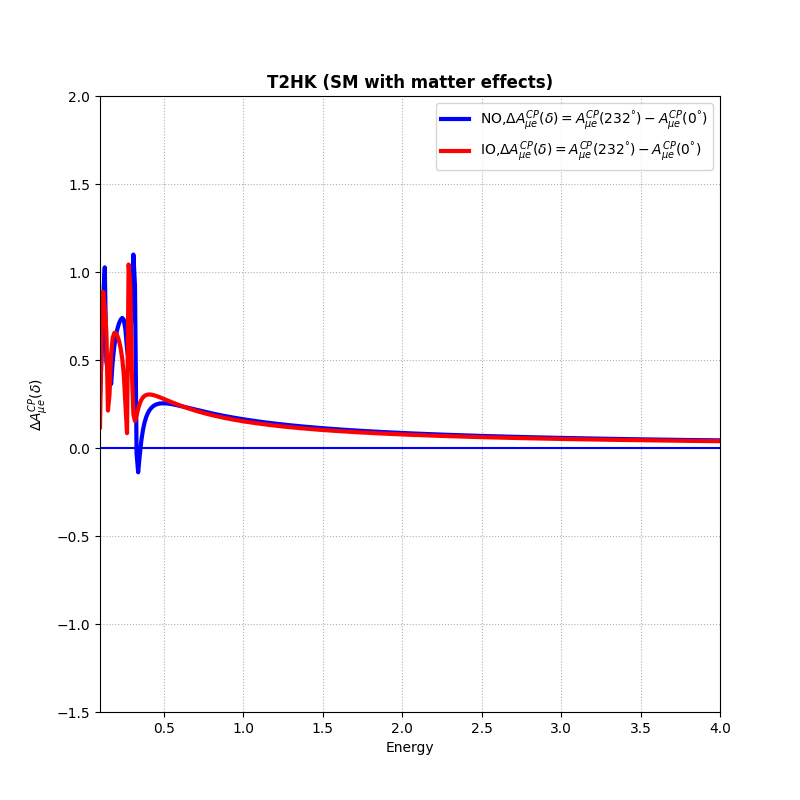}
\endminipage

\minipage{0.45\textwidth}
  \includegraphics[width=8.0cm,height=8.0cm]{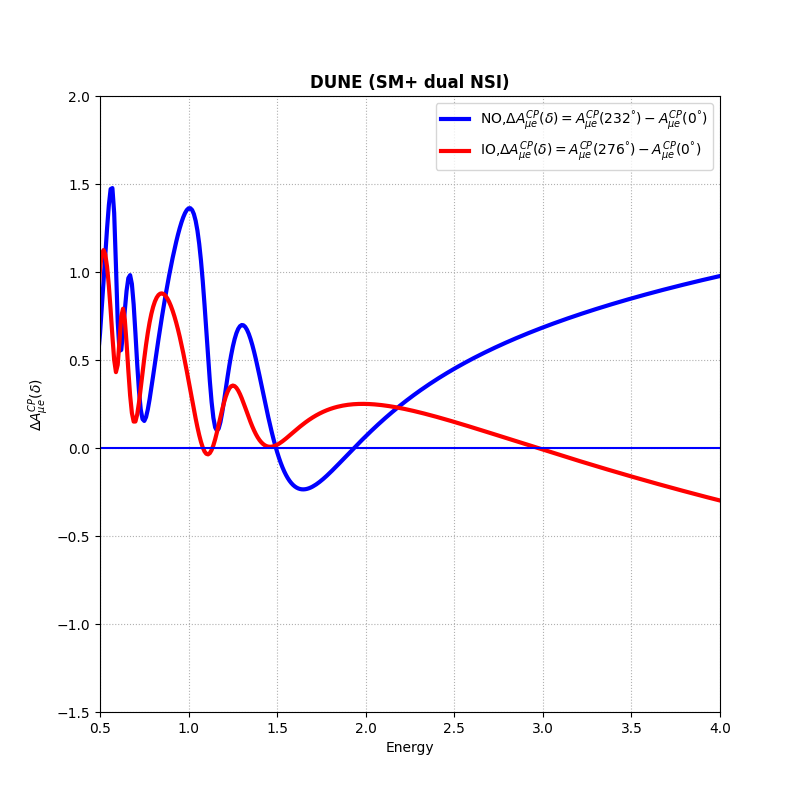}
\endminipage\hfill
\minipage{0.45\textwidth}
  \includegraphics[width=8.0cm,height=8.0cm]{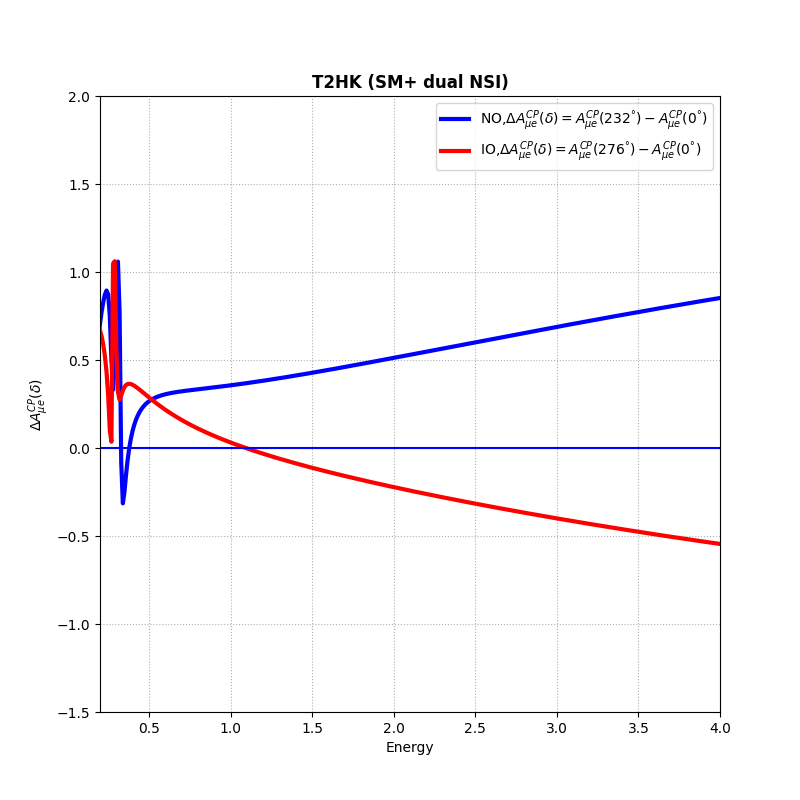}
\endminipage
\caption{Observable $\Delta A_{\mu e}$ versus energy [in GeV] in the case of SM with matter effects (top) and in the presence of dual NSI arising from the $\epsilon_{e\mu}$ and $\epsilon_{e\tau}$ sector simultaneously, in case of DUNE (left) and T2HK (right) experimental setup.}
\end{figure}

\section{Summary and Conclusions}

CP violation in the lepton sector may help us to understand the matter-antimatter asymmetry. Also, in this context, new physics beyond the Standard Model can add new sources of CP violation. Here, in addition to the standard model, we consider that the new physics beyond the standard model is in the form of non-standard neutrino interactions. Needless to mention that the current and future neutrino experiments are likely to provide us with evidence of the CP violation in the neutrino sector. In this analysis, we have utilised roughly 3 years of GLoBES simulated data corresponding to DUNE and T2HK to investigate the CP violation phenomena in vacuum, matter, and in the presence of dual NSI arising from $\epsilon_{e\mu}$ and $\epsilon_{e\tau}$ sectors, simultaneously. Here, we studied the CP asymmetries by taking inputs from ongoing long-baseline neutrino experiments T2K and NO$\nu$A and the standard neutrino parameters from the global fits. Moreover, we have investigated the change in neutrino and anti-neutrino oscillation probabilities in the presence of vacuum, matter, and dual NSI couplings for both NO and IO. Along with the oscillation probability plots, we have also utilised the observable $A_{CP}$ to show a clear distinction between the mass hierarchies in the case of the DUNE experimental setup, whereas in the case of T2HK, there appears to be no such indication. In the case of DUNE, from Figures 5, 6, 13, and 14, we see a clear difference between the mass hierarchies in the presence of matter effects and dual NSI couplings. Furthermore, measuring the observable $\Delta A_{\mu e}$, one can see that both DUNE and T2HK may not be able to distinguish between the mass orderings at their respective energy windows, but for DUNE above 3 GeV and for T2HK beyond 1 GeV energy, a clear distinction between the mass hierarchies are possible. To conclude, with roughly 3 years of simulated data of DUNE and T2HK, one can see the difference in mass hierarchies using the observable $A_{CP}$. The upcoming neutrino experiments era of DUNE and T2HK will provide a better understanding of the neutrino sector and help us to decode the CP asymmetries as well as the long-standing puzzle of neutrino mass hierarchies.

\acknowledgments
The authors would like to thank DST, Govt. of India, for the financial support.

\bibliography{reference.bib}
\bibliographystyle{JHEP}



\end{document}